\documentclass[sigconf]{acmart}
\pdfoutput=1
\clearpage{}\usepackage[utf8]{inputenc} 

\usepackage{graphicx}

\usepackage{xspace}
\usepackage{enumerate}
\usepackage{subfigure}
\usepackage{url}
\usepackage{lipsum}
\usepackage[detect-weight=true,detect-family=true]{siunitx}

\newcommand{\etal}{et al.\xspace}
\newcommand{\ie}{i.e.,\xspace}
\newcommand{\eg}{e.g.,\xspace}

\usepackage[]{todonotes}

\usepackage{cleveref}

\usepackage{amsthm}
\theoremstyle{definition}
\newtheorem{researchquestion}{RQ}
\crefname{researchquestion}{RQ}{RQs}

\newcommand{\numVenues}{11\xspace}    \newcommand{\numPapersIn}{\num{9621}\xspace}  \newcommand{\numShortPapers}{\num{4141}\xspace}  \newcommand{\numNonPrimaryPapers}{\num{28}\xspace}  \newcommand{\numPrimaryPapers}{\num{5452}\xspace}  \newcommand{\numNonParsablePapers}{\num{79}\xspace}  \newcommand{\numParsablePapers}{\num{5373}\xspace}    \newcommand{\numPapersYesMsr}{\num{3367}\xspace}  \newcommand{\pctPapersYesMsr}{\num{62}\%\xspace}  \newcommand{\numYears}{16\xspace}  \newcommand{\yearsRangeBegin}{2004\xspace}  \newcommand{\yearsRangeEnd}{2020\xspace}  \newcommand{\yearsRange}{\yearsRangeBegin--\yearsRangeEnd}  \clearpage{}
\copyrightyear{2022}
\acmYear{2022}
\setcopyright{acmlicensed}
\acmConference[ESEM '22]{ACM / IEEE International Symposium on Empirical Software Engineering and Measurement (ESEM)}{September 19--23, 2022}{Helsinki, Finland}
\acmBooktitle{ACM / IEEE International Symposium on Empirical Software Engineering and Measurement (ESEM) (ESEM '22), September 19--23, 2022, Helsinki, Finland}
\acmPrice{15.00}
\acmDOI{10.1145/3544902.3546239}
\acmISBN{978-1-4503-9427-7/22/09}

 \title [Software Artifact Mining in Software Engineering Conferences: A Meta-Analysis]{Software Artifact Mining in Software Engineering Conferences:\\ A Meta-Analysis}

\begin{abstract}
  \textbf{Background:} Software development results in the production of various types of artifacts: source code, version control system metadata, bug reports, mailing list conversations, test data, etc. Empirical software engineering (ESE) has thrived mining those artifacts to uncover the inner workings of software development and improve its practices. But \emph{which artifacts} are studied in the field is a moving target, which we study empirically in this paper.

  \textbf{Aims:} We quantitatively characterize the most frequently mined and co-mined software artifacts in ESE research and the research purposes they support.

  \textbf{Method:} We conduct a meta-analysis of artifact mining studies published in \numVenues top conferences in ESE, for a total of \numPapersIn papers. We use natural language processing (NLP) techniques to characterize the \emph{types of software artifacts} that are most often mined and their evolution over a \numYears-year period (\yearsRange). We analyze the \emph{combinations} of artifact types that are most often mined \emph{together}, as well as the relationship between \emph{study purposes} and mined artifacts.

  \textbf{Results:} We find that: (1) mining happens in the vast majority of analyzed papers, (2) source code and test data are the most mined artifacts, (3) there is an increasing interest in mining novel artifacts, together with source code, (4) researchers are most interested in the evaluation of software systems and use all possible empirical signals to support that goal.

\end{abstract}



 \author{Zeinab Abou Khalil}
\email{zeinab.abou-khalil@inria.fr}
\orcid{0000-0003-0725-1024}
\affiliation{\institution{Inria}
  \city{Paris}
  \country{France}
}

\author{Stefano Zacchiroli}
\email{stefano.zacchiroli@telecom-paris.fr}
\orcid{0000-0002-4576-136X}
\affiliation{\institution{LTCI, Télécom Paris, Institut Polytechnique de Paris}
  \city{Paris}
  \country{France}
}

\begin{document}
\maketitle

\section{Introduction}
\label{sec:intro}

Software development is a human activity that results in the production of
different types of \emph{software artifacts}. While \emph{source code} is the
most common artifact we tend to think of, other results of software production
include: binary code, version control system (VCS) metadata, bug reports,
developer conversations happening via several media (mailing lists, forums,
Q\&A websites, chats), code reviews, test data, and documentation (design-,
developer- and user-oriented).

Empirical Software Engineering (ESE)~\cite{shull2008eseguide, felderer2020esebook}
research in general, and even more so Mining Software
Repository~\cite{nagappan2009msr, hassan2008road} (MSR) specifically, have
analyzed software artifacts in increasingly large quantities over the last
decades, as a way to understand and improve software development practices.

\emph{Which} artifacts are studied in the field is however a moving target,
affected by factors like research trends, data availability, and increased
availability of large-scale datasets and analysis platforms~\cite{FLOSSmole,
  dyer2013boa, swh-msr2019-dataset, mockus2019woc}. This shift over time has
led researchers in the field to regularly conduct ``introspective''
studies~\cite{kagdi2007survey, robles2010replicating, demeyer2013happy,
  hemmati2013msr, amann2013software, farias2016msrmapping,
  menzies2018swetrends, kotti2020msrdatapapers} that use
meta-research~\cite{ioannidis2010metaresearch} techniques to analyze published
ESE scientific papers and report back to the community.

\paragraph{Contributions}

The present work fits the tradition of (automated)
meta-analyses~\cite{felizardo2020automating} on empirical software engineering
papers, addressing and reporting back to the community on the under-explored
angle of which software artifacts are mined, their co-occurrence in studies,
their relationship with study purposes, and the evolution of their mining over
time. Specifically, we address the following research questions:

\begin{researchquestion}
  What are the \textbf{most frequently mined software artifacts} in ESE
  research, and how has their popularity \emph{evolved over time}?
\label{rq:mined-artifacts}
\end{researchquestion}

\begin{researchquestion}
  What are the \textbf{combinations of software artifacts} that are frequently
  mined together in ESE research?
\label{rq:artifact-combinations}
\end{researchquestion}

\begin{researchquestion}
  What are the most popular \textbf{research purposes} of studies that mine
  software artifacts in ESE research, and how do they \textbf{relate to the
    type of mined artifacts}? (Intuitively: which software artifacts support
  which research purposes in the field?)
\label{rq:artifact-purposes}
\end{researchquestion}

By answering these research questions, we aim to provide a comprehensive view of
how, how much, and why software artifacts are used in ESE research, with a
quantitative and longitudinal (over time) angle. Doing so will not only inform
the community about the evolution of research trends but also guide
\emph{research policy decisions}. In particular, the knowledge that more, or
simply other, types of artifacts are in high demand for ESE research can
motivate fellow scholars to produce needed and hence impactful open datasets;
the knowledge that specific types of artifacts are used \emph{together} can
help in producing datasets that are mutually consistent across different
artifact types, reducing threats to validity due to the use of inconsistent
datasets; realizing that highly used software artifacts types are not being
long-term archived can drive digital preservation
initiatives~\cite{swhipres2017} that aim to support study
repeatability/reproducibility/replicability~\cite{boisvert2016acmbadges} to
focus their efforts on them.

To answer the stated research questions, we mine the textual content of
\numPapersIn from \numVenues top conferences in (empirical) software
engineering, covering a period of \numYears years (\yearsRange). We use NLP
techniques to identify frequently mined software artifacts
and map papers to them. We then study the evolution of detected artifact
mentions over time, the co-occurrence of artifacts types in papers, and the
relationship between paper purposes (also mined from paper texts) and artifact
types.

\paragraph{Paper structure}

A comparison with related work is conducted next in \Cref{sec:relatedwork}.
\Cref{sec:method} describes our experimental methodology. \Cref{sec:result}
presents experimental results, breaking them down by research question. We
discuss the implications of our findings in \Cref{sec:discussion} and threats to
their validity in \Cref{sec:threats}. \Cref{sec:conclusion} summarizes the
paper and outlines directions for future work.

\paragraph{Data availability}

A complete replication package for this paper is available from
Zenodo~\cite{replication-package}.

 \section{Related work}
\label{sec:relatedwork}

Demeyer \etal~\cite{demeyer2013happy} mined the complete corpus of MSR conference papers at the time (2004--2012) using n-gram analysis, to investigate how the research field on mining software repositories had evolved.
They focused on: (i) trendy (and outdated) research topics, (ii) most (and least) frequently cited cases, (iii) popular/emerging mining infrastructures, and (iv) software engineering state-of-the-practice. 
They considered only papers published at MSR, so their results cannot be generalized to other venues.
Our work addresses this problem by covering a much larger set of venues.
They used \texttt{pdftotext},\footnote{\url{https://www.xpdfreader.com/pdftotext-man.html}, accessed 2022-04-27} for text extraction, a tool that introduces artifacts in the extracted text, that the authors had to clean up manually.
The tool we used, CERMINE, has been shown to be more reliable~\cite{tkaczyk2018machine}.
Also, they only excluded bibliography sections, whereas we have verified that other sections (e.g., related work and appendix), can also skew the result; we excluded them from analysis too.
Finally, as no link between n-grams and papers was kept in the study, it was not possible to explain outliers in occurrence frequencies.

Novais \etal~\cite{novais2013software} conducted a systematic study of software evolution visualization technologies over 125 papers. 
They studied the types of data used to visualize and analyze software evolution.
They found that there is no study combining “BTS data” and “Source Code”, and that “SCM data” is the key data source for software evolution visualization.
Their results are specific to software evolution visualization, whereas we conduct a broader analysis covering software artifact mining across top ESE venues.

Farias \etal~\cite{farias2016msrmapping} performed a systematic mapping study on 107 papers published over 5 editions of the MSR conference (2010--2014).
They manually investigated papers, collecting data about software analysis goals (purpose, focus, and object of the analysis), data sources, evaluation methods, tools, and how the field is evolving. They found that “comprehension of defects” and “code” were, respectively, was the most common purpose and analysis object (\ie artifact).
They defined a taxonomy organizing artifacts into structured (\eg source code) and unstructured (\eg mailing lists) ones. They found that structured artifacts tend to be more explored than unstructured ones, but that the number of approaches using unstructured ones had been increasing over the last three years (at the time).
Other software engineering artifacts were starting to come into use at the time, such as comments and emails.
They were being analyzed either alone or together with metrics extracted from structured data sources to understand quality issues in software projects.
Their study is limited to 5 editions of MSR, which are now more than 5 years old.
Our study is more general in terms of both venues and time period.
We adopt their taxonomy of study purposes to answer \Cref{rq:artifact-purposes}.

Amann \etal~\cite{amann2013software} reviewed studies published at top software engineering conferences to describe the current (in 2013) state of the art, trends in mined artifacts, pursued goals and study reproducibility.
For the specific purpose of identifying mined artifacts they only considered papers published at the MSR conference.
They identified 15 distinct artifact ``sources'' including: CVS, git, mercurial, GitHub, SVN, jazz, bug, commit, patch, message, StackOverflow, email, Twitter, blog, and tutorials. 
Similar to~\cite{demeyer2013happy}, authors included only MSR in their analysis and excluded only paper bibliographies from their analyses.
Also, they focus their analysis on the top 10 terms in papers and use product names (\eg Mercurial) rather than artifact types.
In the present study, we investigate artifacts mining in a much larger body of conferences, longer period, and additionally focus on the co-occurrence of artifact types and study purposes to inform data policy decisions.

Hemmati \etal~\cite{hemmati2013msr} analyzed 117 full papers published at MSR between 2004 and 2012.
They extracted 268 comments from these papers, categorized them using a grounded theory methodology, and extracted high-level research themes.
They codified a set of guidelines, tips, and recommendations, as well as a set of best practices, which can be used and updated continuously as the MSR community matures and advances.

Vasilescu \etal~\cite{vasilescu2013sweconf} curated a dataset of 11 well-established software engineering conferences containing historical data about accepted papers, program committee members, and the number of submissions for the 1994--2012 period.
The dataset is intended to assist steering committees or program committee chairs in their selection process (\eg the change in PC members), help potential authors decide where to submit their work to, and study the number of conference newcomers over time.
They used DBLP records to retrieve paper data and extracted PC members and the number of submissions from event websites and online proceedings.

Kotti \etal~\cite{kotti2020msrdatapapers} conducted a study of data papers published at MSR between 2005 and 2018 to determine how often (frequency), by whom (user), and for what purpose researchers reuse associated artifacts.
They found that 65\% of data papers have been used in other studies, but they are cited less often than technical papers at the same conference.
Their findings highlight that data papers provide useful foundations for subsequent studies.

 \section{Methodology}
\label{sec:method}

\begin{figure*}
  \centering
  \includegraphics[width=0.7\textwidth]{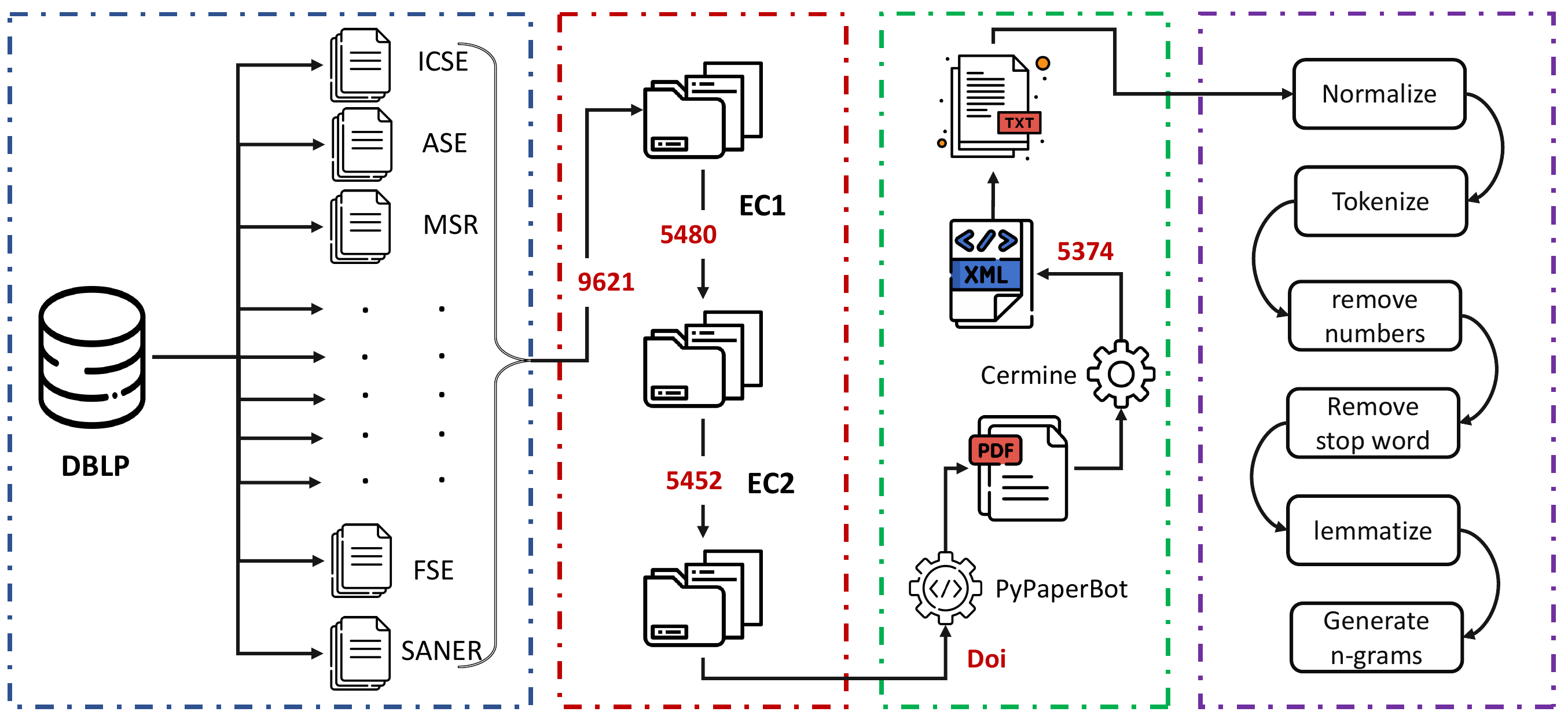}
  \caption{Overview of the methodology as a sequence of the following steps:
    venue selection, paper retrieval and filtering, text extraction, text
    cleanup, textual analysis.}
  \label{fig:method}
\end{figure*}

\Cref{fig:method} gives an overview of our data collection and analysis
approach. We detail each step in the remainder of this section.

\subsection{Venue selection}

Novel results in software engineering tend to be first published in scientific
\emph{conferences} and later (and not always) consolidated in \emph{journal}
articles. Consistently with previous work~\cite{amann2013software,
  vasilescu2013sweconf}, we focus in this paper on conferences, which we posit
to closely capture research trends in ESE at a given point in time.

\begin{table*}
  \centering
    \caption{Selected software engineering conferences and periods, with number
    of papers considered for each (before any filtering and venue merging).}
  \begin{tabular}{clcr} 
    \textbf{Acronym} & \multicolumn{1}{c}{\textbf{Full name}} & \textbf{Period} & \multicolumn{1}{c}{\textbf{Papers}} \\
    \hline
    ASE & IEEE/ACM International Conference on Automated Software Engineering & 2004--2020 & 1702 \\
    ESEC/FSE &  ACM SIGSOFT Symposium on the Foundations of Software Engineering & 2004--2020  & 1503 \\
    ICPC &  IEEE International Conference on Program Comprehension & 2006--2020 & 635 \\
    ICSE  & International Conference on Software Engineering & 2004--2020 & 2082 \\
    ICSM  & IEEE International Conference on Software Maintenance & 2004--2013 & 810 \\
    ICSME & International Conference on Software Maintenance and Evolution & 2014--2020 & 636 \\
    MSR  & Working Conference on Mining Software Repositories & 2004--2020 & 838 \\
    SCAM  & International Working Conference on Source Code Analysis \& Manipulation  & 2004--2019 & 407 \\
    WCRE &  Working Conference on Reverse Engineering & 2004--2013 & 464 \\
    CSMR &  IEEE Conference on Software Maintenance, Reengineering, and Reverse Engineering & 2014--2014 & 71 \\
    SANER & IEEE International Conference on Software Analysis, Evolution and Reengineering & 2015--2019 & 474 \\
    \hline
    \textbf{Total} & & \textbf{\numYears years} & \textbf{\numPapersIn} \\
  \end{tabular}

  \label{tab:venues}
\end{table*}

The most representative conference for research based on software artifacts is
Mining Software Repositories (MSR), established in 2004 as the primary
conference for mining-based empirical software engineering, and has successfully
continued ever since. But mining-based ESE studies are also published in other
reputable software engineering conferences; so one cannot \emph{only} consider
MSR papers. The list of conferences we have analyzed in this study is given in
\Cref{tab:venues}, with details about the considered editions and the number of
papers initially obtained from each of them.

To settle on this list of conferences, we started from the list of venues used
in previous work~\cite{vasilescu2013sweconf, menzies2018swetrends} and retained
only conferences that explicitly welcome and/or frequently publish empirical
studies based on software artifact mining.

\numVenues top conferences in (empirical) software engineering were retained.
Some have a broad scope (e.g., ICSE), while others are focused on specific
subdomains of software engineering (e.g., MSR, ICSME). Regarding selected
conference \emph{years}, we started from papers published in 2004---the year MSR was
first held as an international workshop, in response to the increasing interest in this field, denoting the beginning of a more
established field---and stopped at 2020 (because 2021 was still incomplete at
the time of data collection). In total, \numPapersIn papers have been
considered over a period spanning \numYears years.

In reading \Cref{tab:venues}, note that some conferences merged and/or changed
name names during the observation period: ICSM renamed to ICSME; WCRE and CSMR
merged into SANER. In the rest of the analysis, we hence merged ICSM and ICSME
papers, as well as WCRE, CSMR and SANER papers (referred to as ``venue
merging'' in the following).

\subsection{Paper filtering and retrieval}

To obtain the list of papers for the selected conferences and years, we used
DBLP~\cite{ley2002dblp}, the reference bibliographic database for computer
science publications. The entire DBLP dataset is released publicly in XML
format; we retrieved the most recent DBLP data dump available at the time of
analysis.\footnote{We retrieved the dataset from
  \url{https://dblp.uni-trier.de/xml/} in April 2021; specifically, we used the
  data dump named \texttt{dblp-2021-04-01}.}\\
As DBLP indexes studies in all fields of computer science, we first selected
papers from venues and editions that match \Cref{tab:venues}, obtaining
\numPapersIn bibliographic records. We then further filtered records applying
the following \emph{exclusion criteria (ECs)} (matching any one of them is
enough for excluding a paper from further analysis):
\begin{enumerate}[\bfseries EC1]

\item The \emph{paper must not be short} (including position papers, data papers, and
  challenge papers, e.g., those corresponding to the yearly MSR mining
  challenge). The rationale for this criterion is that full papers present more
  mature and established results, rather than exploratory ideas that might not
  bear fruit. As such, full papers better capture the state of the
  field~\cite{farias2016msrmapping}.

  To implement this criterion, we discarded papers shorter than 6 pages,
  according to DBLP metadata.

\item The paper must describe \emph{primary research} rather than secondary or
  tertiary one. (Secondary research is based on published data and
  information gathered from previous studies. Tertiary research is
  meta-research based on secondary studies, such as systematic reviews of
  secondary studies.) The rationale for this choice is that secondary and tertiary
  studies (\eg systematic literature reviews, systematic mappings, and
  literature surveys) would over-represent research trends that are already
  popular in the field.

  To implement this criterion, we exclude papers that contain strings like the
  following in their titles: ``literature review'', ``systematic mapping'',
  ``systematic review'', ``replication study'', ``survey'', ``replicating''
  (see replication package~\cite{replication-package} for
  details).

\end{enumerate}

\begin{figure}
  \centering
  \includegraphics[width=0.9\columnwidth]{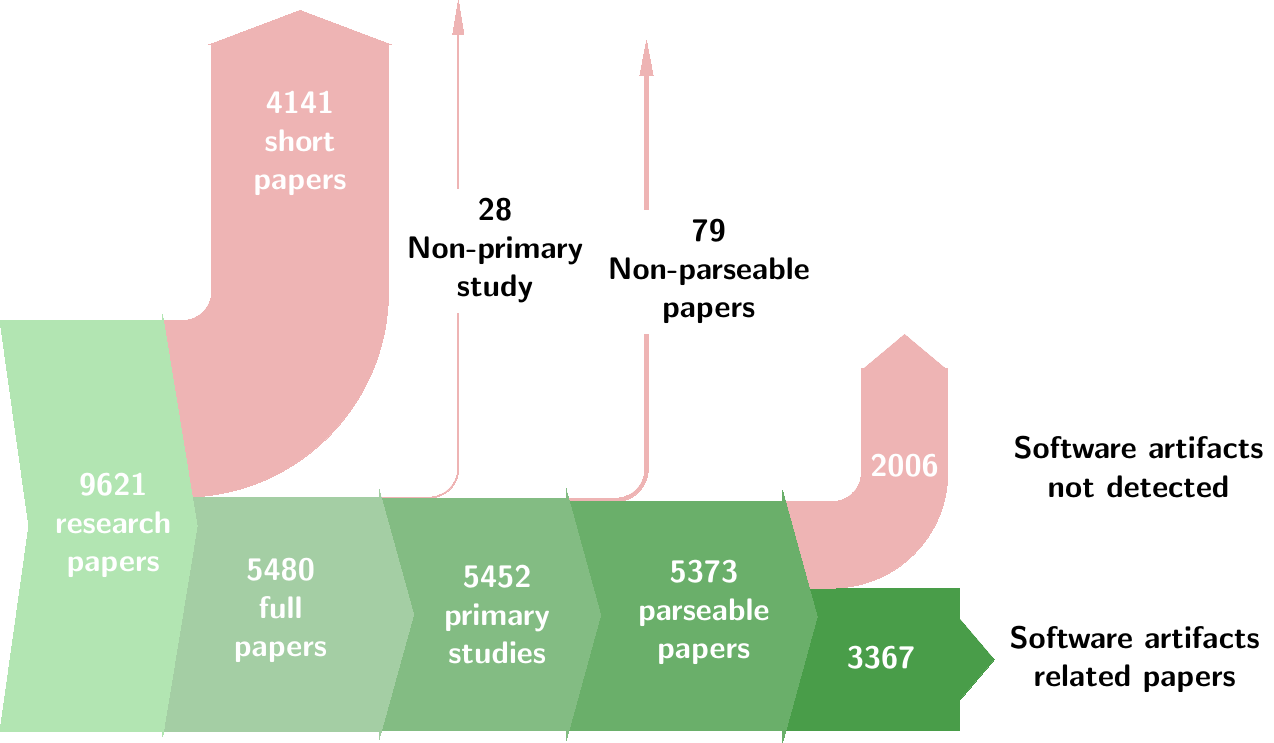}
  \caption{Paper filtering from the initial dataset of \Cref{tab:venues} to the
    final determination of which study deals with software artifact mining.}
  \label{fig:paper-filtering}
\end{figure}

\noindent
The impact of each filtering step is shown as a Sankey diagram in the leftmost
part of \Cref{fig:paper-filtering}. We started from \numPapersIn papers,
\numShortPapers of which were short, and \numNonPrimaryPapers were non-primary
studies. After filtering, \numPrimaryPapers remained.

For all remaining papers, we retrieved DOIs (Digital Object Identifiers) from
either DBLP records (for the most part) or using the Crossref search
engine\footnote{\url{https://search.crossref.org/}, accessed 2022-01-18} (for
about 100 papers, for which DBLP lacked DOI information). We then retrieved digital copies, in PDF format, of selected papers using the
PyPaperBot paper retrieval
tool.\footnote{\url{https://github.com/ferru97/PyPaperBot}, accessed
  2022-01-06. We slightly modified the tool to store PDFs locally using DOIs as
  filenames. Our modified version is included in the replication
  package~\cite{replication-package}.}

\subsection{Text extraction}

Once we obtained all papers in PDF format, we converted them to plain text for
ease of further processing.

Several open source tools exist for this task.  Tkaczyk
\etal~\cite{tkaczyk2018machine} conducted an evaluation of 10 such tools,
showing that GROBID (GeneRation Of BIbliographic
Data)\footnote{\url{https://github.com/kermitt2/grobid}, accessed 2022-01-18}
and CERMINE~\cite{tkaczyk2015cermine} (Content ExtRactor and MINEr) perform best.
Among the two, CERMINE is a comprehensive tool for the automatic extraction of
paper metadata and content. Most importantly, CERMINE returns the paper's full
text \emph{structured} in sections and subsections. As in the following, we need
to discriminate paper text based on the section it appears in, we used CERMINE
for PDF-to-text conversion.

As shown in \Cref{fig:paper-filtering}, we could not parse
\numNonParsablePapers (out of \numPrimaryPapers remaining thus far), due to few
PDF files containing pages encoded as bitmaps rather than structured
content. After removing them, \numParsablePapers remained for textual analysis.

\subsection{Textual analysis}

We performed a series of preparation and cleaning steps on the XML files
produced by CERMINE before further analysis.

First, we retrieved the top used section headers from all papers for manual
inspection. Then we eliminated from the extracted paper texts all content
belonging to sections like References, Related Work, Background,
Acknowledgement(s), Bibliography, Future Work, and Limitations (we used various
variants of this section names, based on the results of popular headers; see
the replication package~\cite{replication-package} for full
details). The rationale for this choice is that these sections are meta
w.r.t.~the main content of the paper; whether a paper is mining software
artifacts or not will be primarily determined by naming those artifacts in
\emph{other} sections than these.

Once we obtained this corpus of relevant text from the papers, we applied
classical NLP (Natural Language Processing) cleanup steps to reduce noise and
improve the corpus quality.
We started by tokenizing: splitting the text stream into words, symbols,
punctuation, and postfixes using the Natural Language Toolkit Python
library.\footnote{\url{https://www.nltk.org/}, accessed 2022-01-07} Then, all
upper case characters were converted to lower case. Next, we removed
non-alphabetic tokens, and removed stop words. Finally, we applied
lemmatization~\cite{kettunen2005stem}, 
in order to use one canonical representation per term (\eg ``codes'' becomes
``code'') and avoid under-counting popular terms that occur in different forms.

Searching for individual words is often insufficient in text search because
abstract terms are often represented as word sequences, such as ``bug report''
or ``source code''~\cite{moro2015business}. To solve this issue, computational
linguistics commonly uses \emph{n-gram analysis}---where an n-gram is a sequence
of $n$ contiguous words---to analyze large document corpora in order to
identify combinations of words that frequently appear
together~\cite{manning1999foundations}. Soper and Turel~\cite{soper2012n}
suggested that using n-gram analysis allows computer scientists and scholars to
gain insights into vast document corpus, such as ours.

We conducted an n-gram analysis on the paper corpus as filtered thus far. We
limited the n-grams analysis to a length of a maximum of 2 (i.e., 1-grams and
2-grams) as longer n-grams are rarely repeated in texts. At the end of the
process, we obtained a set of 1,2-grams with the number of occurrences for each
paper.

We transformed all n-grams with associated origin information (i.e., the papers
they occur in) to a list of records and saved them to a CSV file. Each record
consist of the following fields: \{ngram, frequency, year, DOI, acronym\}. For
example \{``bug report'', 12, 2018, ``10.1145/xxx.xxx'', ``MSR''\} means: the
n-gram ``bug report'' occurred 12 times in the (filtered) text of a paper
published at MSR 2018 whose DOI is ``10.1145/xxx.xxx''.

\begin{table}

  \caption{Excerpt of the top 2-grams extracted from the paper corpus, with
    software artifact types they are associated to.}
  \small
  \begin{tabular}{clcl}
    \toprule
    \textbf{Rank} & \textbf{N-gram} & \textbf{N.~of papers} & \textbf{Artifact type} \\
    \midrule
    0   &             (source, code) &  1926 & source code\\
    1   &               (test, case) &   771 & test data\\
    2   &              (test, suite) &   494 & test data\\
    3   &               (line, code) &   391 & source code\\
    4   &              (bug, report) &   350 & bug data\\
    6   &             (code, change) &   293 & source code\\
    7   &                 (bug, fix) &   262 & bug data\\
13  &            (code, snippet) &   199 & source code\\
    15  &           (code, fragment) &   194 & source code\\
    16  &             (source, file) &   193 & source code\\
    22  &              (code, clone) &   167 & source code\\
    23  &               (unit, test) &   166 & test data\\
35  &             (code, review) &   124 & code review\\
41  &          (commit, message) &   113 & commit metadata\\
    43  &               (code, file) &   110 & source code\\
    45  &               (test, data) &   108 & test data\\
    46  &          (test, execution) &   104 & test data\\
49  &          (stack, overflow) &   102 & forum\\
    50  &            (pull, request) &   102 & commit metadata\\
54  &              (code, smell) &    99 & source code\\
    55  &      (fault, localization) &    99 & bug data\\
81  &       (defect, prediction) &    84 & bug data\\
106 &               (mail, list) &    75 & mail data\\
    110 &               (java, code) &    74 & source code\\
116 &           (class, diagram) &    72 & uml diagram\\
144 &             (change, code) &    67 & source code\\
    \bottomrule
  \end{tabular} 
  \label{tab:ngram-taxonomy}
\end{table}

\subsection{Detection of mined artifacts}

We manually inspected the top-150 1- and 2-grams mined from the paper corpus and partitioned them into a taxonomy of 8 classes of software artifacts, namely: (1) bug data, (2) source code, (3) mail data (e.g., mailing list discussions), (4) code review, (5) commit metadata, (6) test data, (7) forum data (encompassing discussions on Web forums, Q\&A websites like StackOverflow, microblogging platforms and Reddit), and (8) UML diagrams (including other diagram languages for describing software systems).
While we applied our own judgment and domain knowledge to come up with this taxonomy, what we obtained is consistent with, and in fact a subset of, the taxonomy developed by Farias \etal~\cite{farias2016msrmapping}.

\Cref{tab:ngram-taxonomy} shows an excerpt of the top 2-grams mined from paper texts together with the associated artifact types.
For the most part, 1-grams were too generic (e.g., ``software'' and ``data'' are the first two 1-grams) to denote the use in papers of specific artifact types, with the exception of ``twitter'' and ``reddit'' which we mapped to ``forum'' (not shown in \Cref{tab:ngram-taxonomy}).
The complete mapping table from n-grams to artifact classes is included in the replication package~\cite{replication-package}.

To automatically associate papers to mined artifact types, we searched paper texts (searching only within the non-excluded sections discussed before) for the n-grams in the n-gram/artifact mapping table.
Papers containing at least 3 occurrences of 2-grams in the mapping table (respectively: at least 4 occurrences of 1-grams) were considered as analyzing the corresponding artifact type.
The use of thresholds is meant to avoid stray mentions of n-grams (e.g., a paper mentioning ``twitter'' only once is more likely to be pointing to a single Twitter profile than mining developer microblogging messages); the difference in thresholds is due to the higher popularity of shorter n-grams in natural language texts.

\subsubsection*{Validation}

To evaluate the accuracy of the n-gram-based detector of artifacts mined in
analyzed papers, we manually inspected 200 randomly selected papers (3.7\% of
parsable papers in the dataset). Authors have read each paper, noting down
which software artifacts (if any) were mined in the described study.

\begin{table}
  \centering
  \caption{Accuracy of the detector of artifacts mined in analyzed papers,
    based on a ground truth of 200 papers (3.7\% of parsable papers) manually
    reviewed by the authors.}
  \label{tab:validate}
  \resizebox{\columnwidth}{!}{\begin{tabular}{lrrrrrrr}
    \toprule
    \textbf{Artifact type}
     & \textbf{TP} & \textbf{FP} & \textbf{TN} & \textbf{FN}
     & \textbf{Accuracy} & \textbf{Precision} & \textbf{Recall} \\
    \midrule
    bug data        &  17 &   1 &  167 &  15 & 0.92  &  0.94 &  0.53 \\
    mail data       &   5 &   2 &  192 &   1 & 0.98  &  0.71 &  0.83 \\
    source code     &  68 &  21 &   88 &  23 & 0.78  &  0.76 &  0.75 \\
    code review     &   2 &   2 &  194 &   2 & 0.98  &  0.50 &  0.50 \\
    forum           &   4 &   6 &  188 &   2 & 0.96  &  0.40 &  0.67 \\
    commit metadata &  10 &   0 &  176 &  14 & 0.93  &  1.00 &  0.42 \\
    test data       &  26 &  17 &  154 &   3 & 0.90  &  0.60 &  0.90 \\
    uml diagram     &   2 &   0 &  197 &   1 & 1.00  &  1.00 &  0.67 \\
    \bottomrule
    \textbf{Average} &    &     &      &     & 0.931 & 0.738 & 0.658
    \\
    \bottomrule
  \end{tabular}
  }
\end{table}

\begin{table*}
  \centering
  \caption{Taxonomy of study purposes (from~\cite{farias2016msrmapping}) and associated n-grams (excerpt).}
  \small
  \begin{tabular}{l p{7.8cm} p{6.9cm}}
    \toprule
    \textbf{Purpose} & \textbf{Description} & \textbf{N-grams (excerpt)}  \\
    \midrule
    Comprehension & Comprehension of software systems, including source code & (program, comprehension), (code, comprehension) ...\\ 
    Prediction &  Prediction of changes in systems properties, including bugs, failures, and quality aspects & ('fault', 'prediction'),
    ('change', 'prediction') ...\\
    Contribution analysis & Analyze developer activities & ('developer', 'contribution')\\
    Quality improvement & Investigate or propose solutions on how to improve software quality & ('quality', 'improvement') \\
    Process improvement & Help to decide how to improve software processes & ('process', 'improvement')\\
    Software reuse  & Help developers to reuse software more effectively & ('software', 'reuse')\\
    Classification  & Classify software artifacts (defects, programming texts, authorship, etc.) & ('bug', 'classification'),
    ('defect', 'classification'),
    ('authorship', 'classification'),
    ('change', 'classification'),
    ('text', 'classification') ...\\
    Localization & Determine the origin of bug and faults & ('fault', 'localization'), ('bug', 'localization') ...\\
    Evaluation & Tools and approaches to evaluate software systems (including human aspects) & ('performance', 'evaluation'),
('qualitative', 'evaluation'),
('developer', 'evaluation') ...\\
    \bottomrule
  \end{tabular}
  \label{tab:purpose-taxonomy}
\end{table*}

\Cref{tab:validate} summarizes the precision and recall obtained for each
artifact type. Each paper can mine different types of artifacts. Precision for
an artifact type hence refers to the proportion of studies that were correctly
assigned to a given artifact type; recall refers to the proportion of studies
that were correctly assigned to a given artifact type, among those that truly
(based on manual inspection) mine artifacts of the given type. On average, both
precision and recall are satisfactory, validating the findings discussed in the
following.

\subsection{Detection of study purposes}

To automatically detect paper purposes, we followed the same approach used for artifact detection, with the following differences.
We adopted the taxonomy of paper purposes from Farias \etal~\cite{farias2016msrmapping}, which was developed to classify MSR study purposes using the approach of Basili \etal~\cite{basili1999building}.
\Cref{tab:purpose-taxonomy} recalls the purpose taxonomy and gives an example of n-grams associated with each of them; the full mapping from n-grams to purposes is included in the replication package~\cite{replication-package}.
To associate papers to purposes, we analyzed top n-grams extracted only from abstracts, as abstracts generally state explicitly what the study purpose is and are (for purpose detection) less noisy than the rest of the paper text.

 \section{Results}
\label{sec:result}

We present in this section our experimental results, answering in order the
research questions stated in \Cref{sec:intro}.

\subsection{\Cref{rq:mined-artifacts}: most popular software artifacts}

This research question aims to establish which software artifacts are mined the most in software engineering conference papers, and how their amounts evolve over time.
We answer this question for the 8 types of artifacts identified in \Cref{tab:ngram-taxonomy}.
By searching for enough occurrences (after threshold verification) of the corresponding n-grams in the relevant sections of \numParsablePapers parsable papers, we found that \numPapersYesMsr (\pctPapersYesMsr of parsable papers) mine software artifacts of one or more of the selected types.
\textbf{Software artifact mining happens in the vast majority, almost 2/3, of the papers} we have analyzed.

\begin{figure}
  \centering
  \includegraphics[width=\columnwidth]{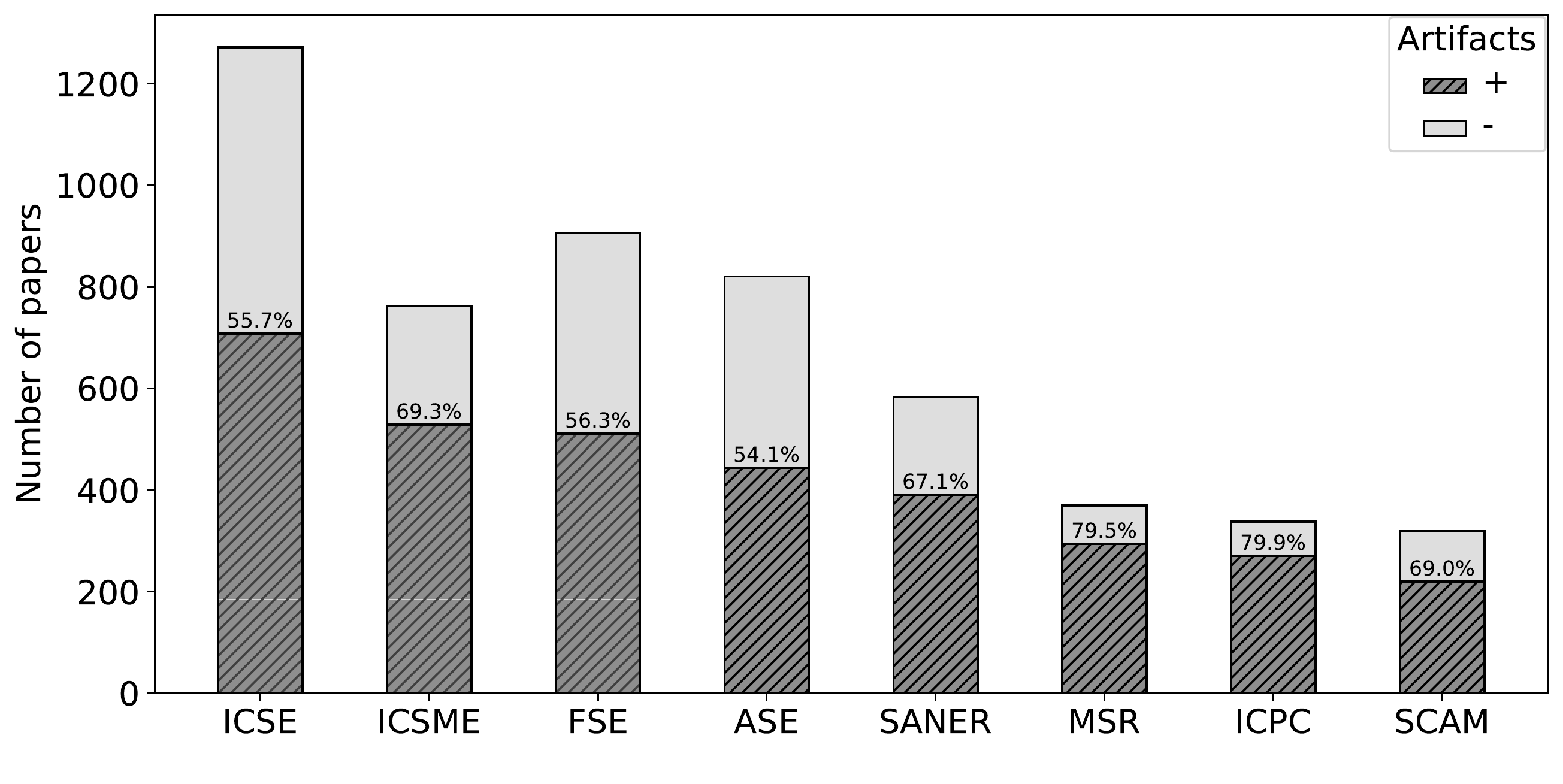}
  \caption{Total number of papers per conference (after venue merging), with
    breakdown of papers detected as mining at least one kind of software
    artifacts (denoted with ``+'') v.~papers detected as not mining any kind of
    artifacts (``-'').}
  \label{fig:Pyear+}
\end{figure}

Different venues exhibit different ratios of software artifact mining, though.
\Cref{fig:Pyear+} provides a breakdown by conference of the total number of papers analyzed together with the ratio of papers detected as mining software artifacts by venue.
Note how more generalist conferences (ICSE, FSE, ASE) contribute a higher number of papers to the corpus, but have a lower ratio of papers mining software artifacts.
Those ratios are high nonetheless, above 50\%; it appears that software artifact mining is a foundational research technique in software engineering, popular also in generalist venues.
The ratios of papers mining software artifacts are even higher in specialized conferences, such as MSR, ICPC, SCAM and, to a lesser extent, ICSME and SANER.

\begin{figure*}
  \centering
  \subfigure[Entire corpus]{\includegraphics[width=0.9\columnwidth]{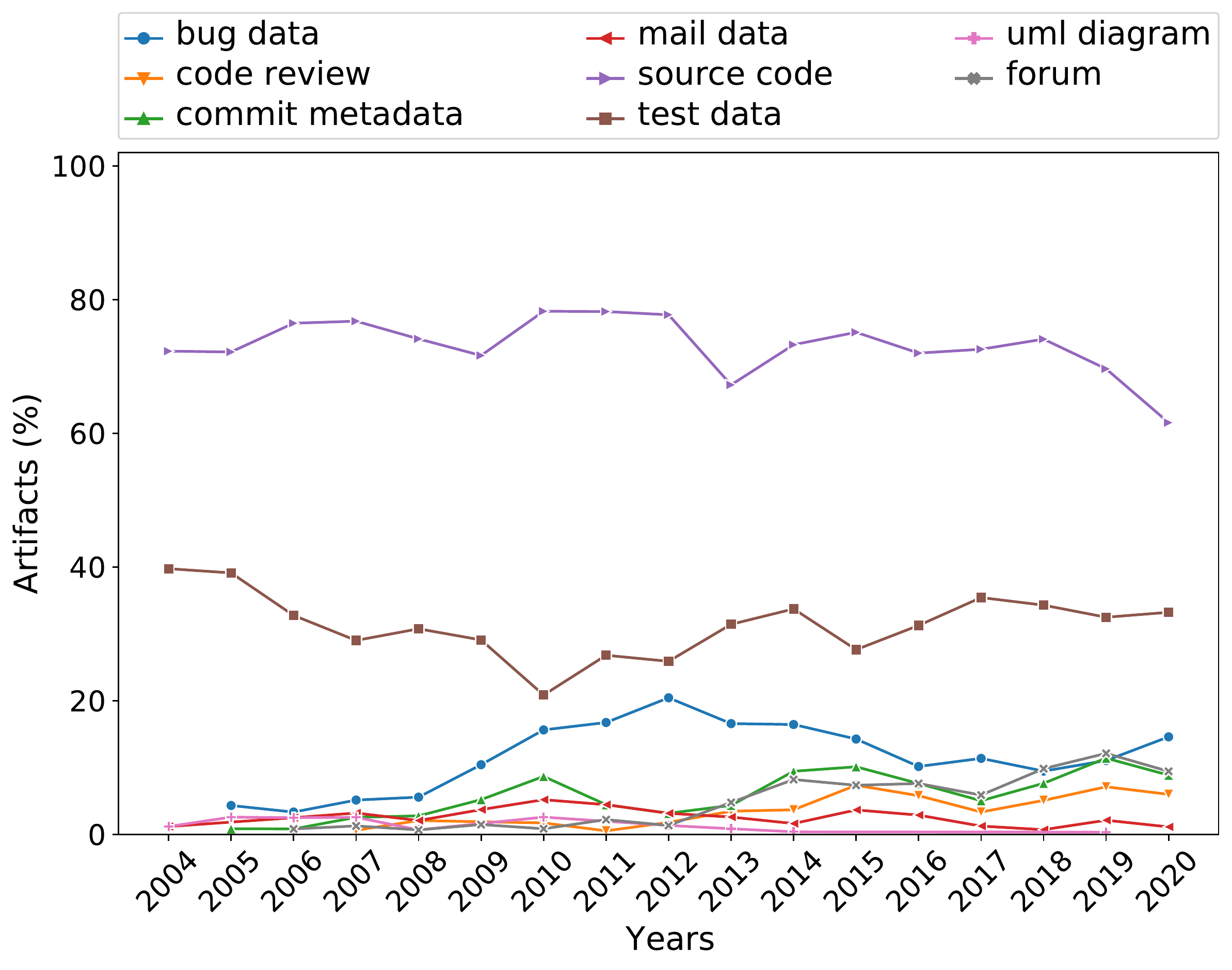}
  \label{fig:artifacts-all}}
  \subfigure[MSR papers]{\includegraphics[width=0.9\columnwidth]{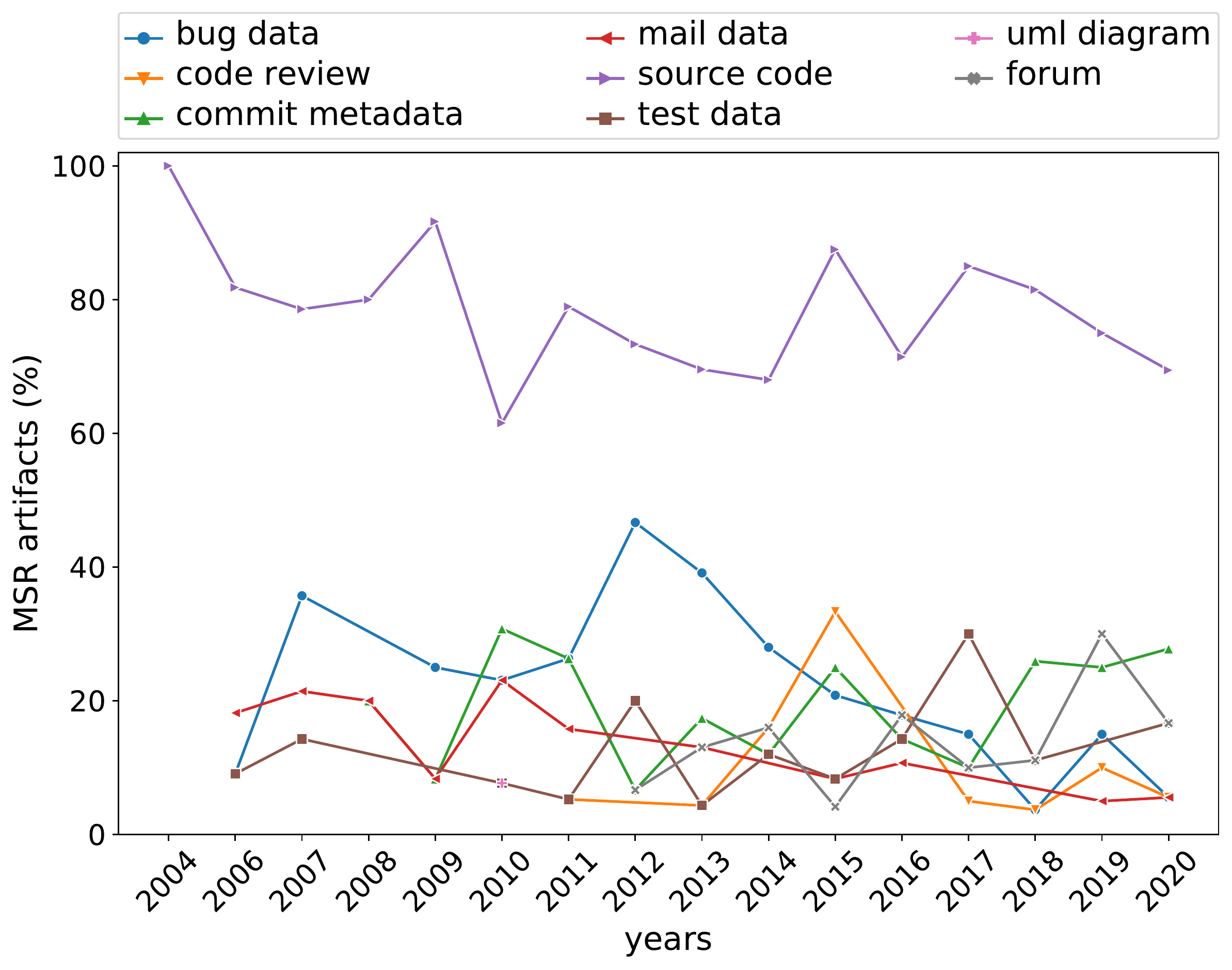}
  \label{fig:artifacts-msr}}
  \subfigure[SCAM papers]{\includegraphics[width=0.9\columnwidth,trim=0 0 0 2.9cm,clip]{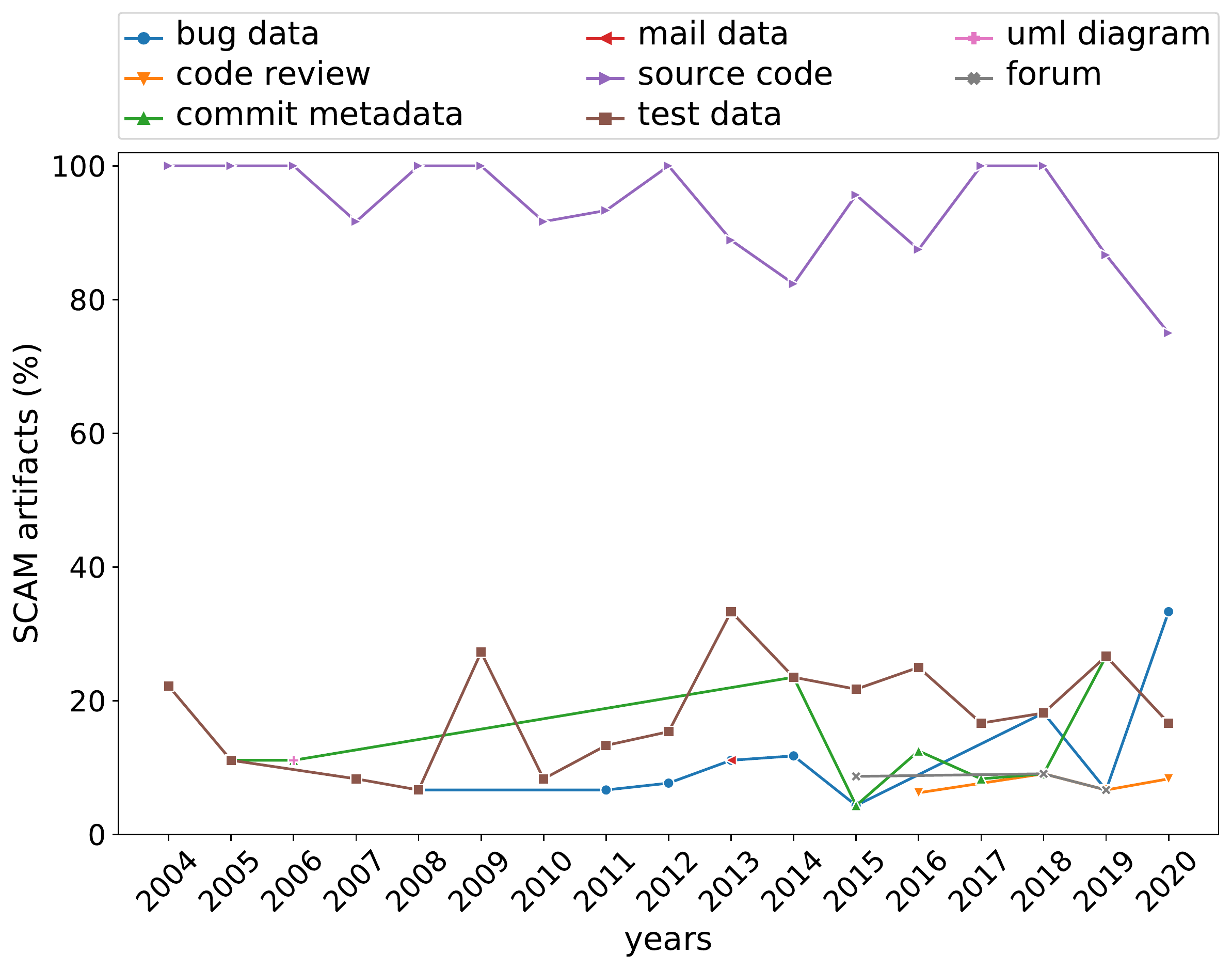}
  \label{fig:artifacts-scam}}
  \subfigure[ICSE papers]{\includegraphics[width=0.9\columnwidth,trim=0 0 0 2.9cm,clip]{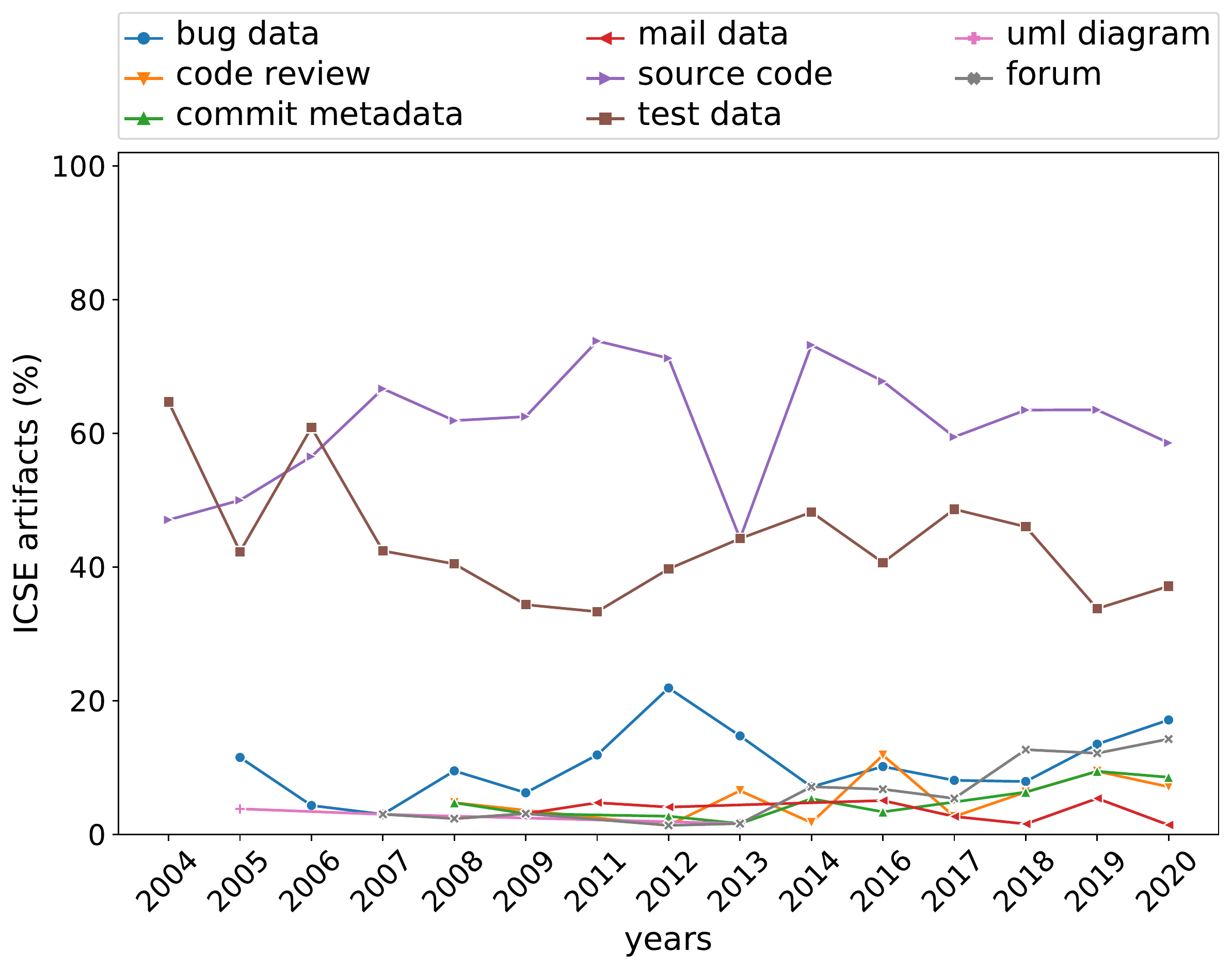}
  \label{fig:artifacts-icse}}
\caption{Percentages of papers mining specific types of software artifacts
    over time, for the entire corpus. Note that, as each paper can mine
    multiple types of artifacts, percentages may add up to more than 100\%.}
\label{fig:artifacts}
\end{figure*}

Regarding the time dimension of \cref{rq:mined-artifacts}, \Cref{fig:artifacts-all} shows the percentages of studies mining each type of artifact, over the years and for the entire corpus.
Note that, as the same paper can mine multiple types of artifacts, percentages do not add up to 100\%.

“Source code” is the most mined software artifact, being ranked first throughout the entire period, with a percentage range fluctuating between 64.3--78.9\%.
It is followed by “test data”, consistently ranked second, within a 24.6--39.7\% range.
\textbf{Source code and test data are the most mined software artifacts} in analyzed studies, and the interest in them by the community has been quite stable over the past \numYears years.
The interest in analyzing “bug data” comes third. It started to increase in 2008, culminated at 19.9\% of yearly papers in 2012, then started decreasing.
The number of ``commit metadata'' studies was negligible before 2005 and increased over time after that.
Most notably, in both 2016 and 2019, the relative interest in commit metadata was either the same or slightly superior to that in bug data.
The number of studies concerned with analyzing unstructured artifacts (\eg mailing list) increased slightly over time starting from 2006, consistently to what Farias \etal~\cite{farias2016msrmapping} observed for a shorter time period, however, it still corresponds to a very small percentage of all analyzed studies.

Given that the MSR conference was launched as a venue dedicated to mining software repositories studies, which is the main theme of our meta-analysis, we zoom into papers published at MSR with \Cref{fig:artifacts-msr}.
Comparing \Cref{fig:artifacts-all,fig:artifacts-msr}, we can see that source code is the dominating software artifact at MSR as well, even more so than in the entire corpus.
However, already starting in 2005, MSR papers exhibit a larger variety of mined artifacts.
Bug data and commit metadata are taking turns in being the second most mined software artifacts.
We also notice in MSR papers higher percentages of unstructured software artifacts being analyzed: mailing list but also code review, which are the second most cited artifact type in MSR 2015 papers. \\
Due to space limitations, in \Cref{fig:artifacts} we only highlight results for selected conferences; results for all conferences individually are included in the paper replication package~\cite{replication-package}.
By comparing conferences, we notice that source code is even more prominent in ICPC and SCAM papers (\Cref{fig:artifacts-scam}), reaching 100\% of papers in multiple editions of each conference.
Conversely, source code is (relatively) less prominent at more foundational software engineering conferences like ICSE (\Cref{fig:artifacts-icse}), FSE, and ASE; it still remains the most referenced artifact type, but in specific years it is overtaken by test data.

\subsection{\Cref{rq:artifact-combinations}: software artifact combinations}

To answer \cref{rq:artifact-combinations} we consider the papers detected as mining more than one type of software artifacts.
28.5\% of the considered papers do so; the rest either do not reference any type of detectable software artifacts or only mention a single artifact type.

\begin{figure}
  \centering
  \includegraphics[width=\columnwidth]{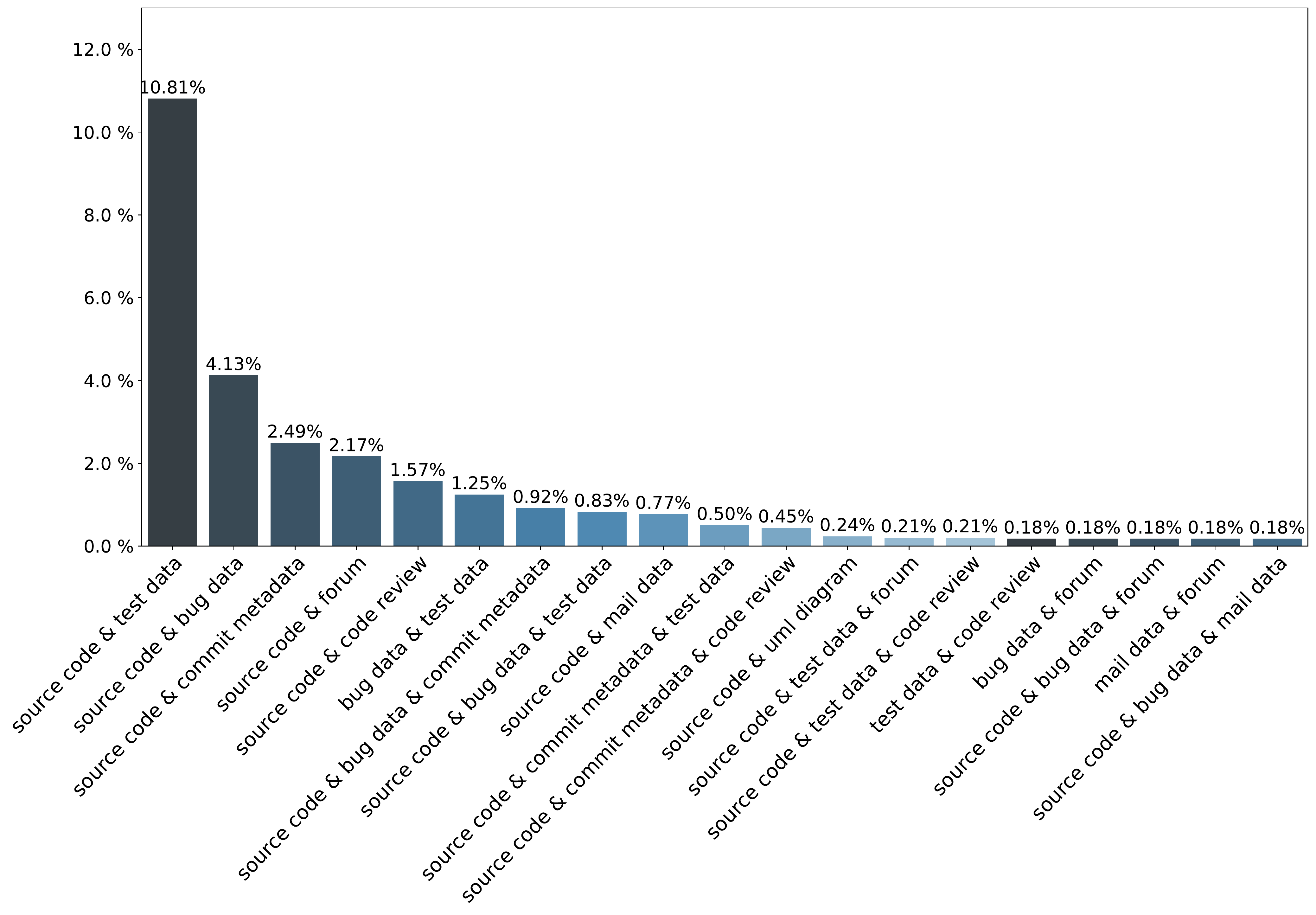}
  \caption{Ratio of the number of papers that mine more than one type of
    software artifacts, by decreasing order of mined artifact combination,
    across all venues.}
  \label{fig:combination}
\end{figure}

\Cref{fig:combination} provides a breakdown of the most popular (top-14) artifact \emph{combinations} found in the paper corps.
The most common combination is of the two types of artifacts which are also the most common ones (individually) that we have identified answering \cref{rq:mined-artifacts}: source code \& test data.
The second combination is source code \& bug data, less than half as popular as the first combination, followed by a more smooth decrease.
After the top-10 combinations, artifact combinations become increasingly more marginal in popularity, with percentages lower than 0.3\%.

Source code is present in most of the top combinations---all except: bug data \& test data, ranked 5th, and test data \& code review, ranked 14th---reinforcing the observation that source code is the most relevant software artifact in empirical software engineering (ESE). Test data, while very popular (6 combinations out of 14) appears to be on more equal footing with bug data (5 combinations) and commit metadata (4 packages) than it appeared to be for \cref{rq:mined-artifacts}.

The topmost combination (ranked 6th) of more than two artifact types mined together is source code \& bug data \& test data, followed by source code \& commit metadata \& test data (8th).
Other popular combinations of at least three artifact types include permutations of these three artifact types and code reviews.
The fact that these four artifact types are also well supported by state-of-the-art social coding platforms like GitHub and GitLab is hardly a coincidence: \textbf{ESE researchers study, for the most part, data that is made accessible by modern coding technology}.
In particular, we observe that social coding platforms have increased the availability of code review and bug data, and that seems to correlate with increased research interest in studying those artifacts.

\begin{figure}
  \centering
  \includegraphics[width=0.7\columnwidth]{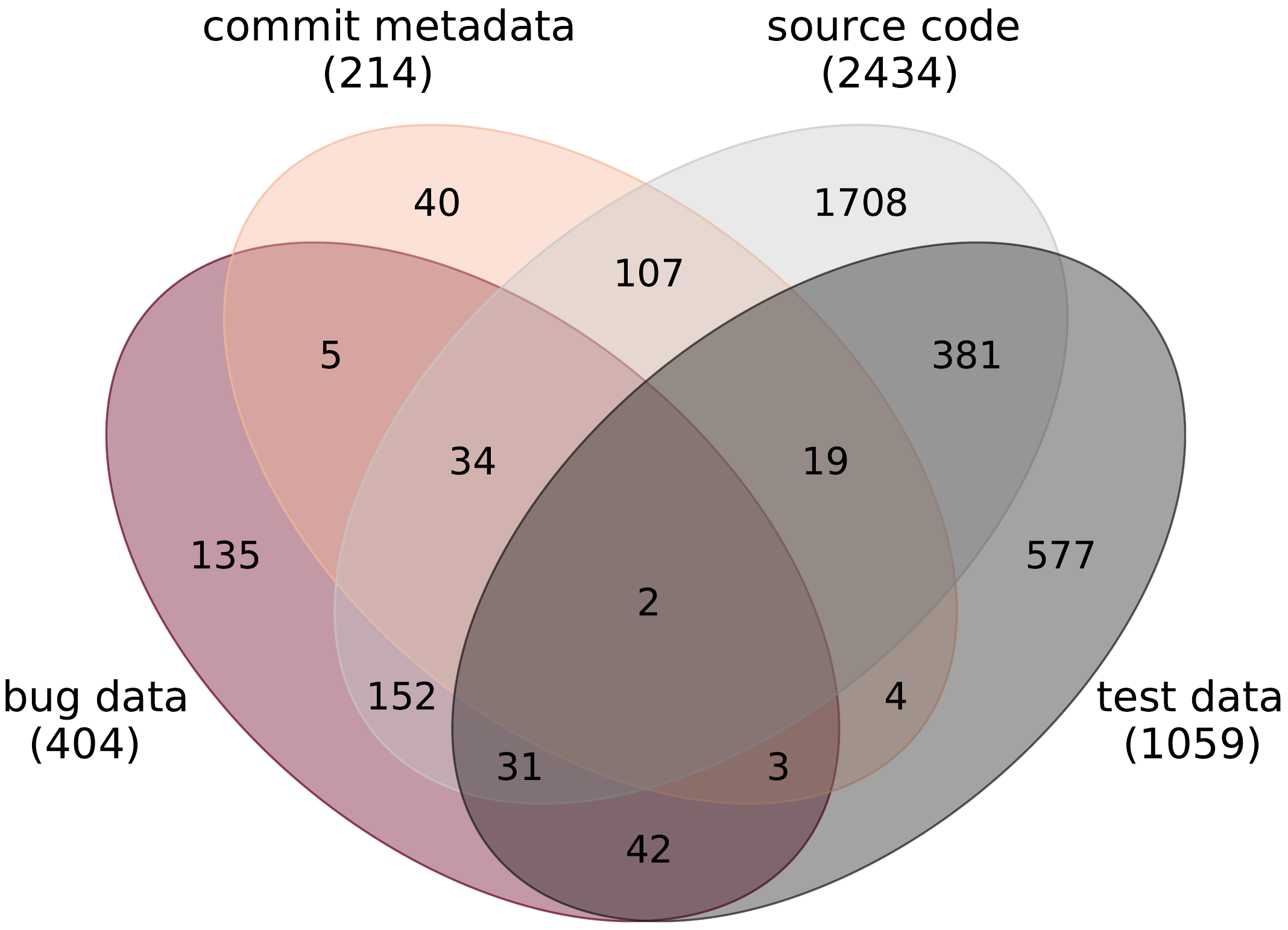}
  \caption{Co-occurrence of the mining of the four most popular artifact types
    (source code, test data, bug data, commit metadata) as sets of papers
    mining any combination of them. Shown as a Venn diagram across all venues.}
  \label{fig:4venn}
\end{figure}

\Cref{fig:4venn} focuses on the four most popular artifact types and their intersections.
It shows a Venn diagram of the studies (each paper is an element in one of the four sets) that references source code, test data, bug data, and/or commit metadata.
Each paper contributes a single set element to the diagram, and papers not referencing any of those artifacts have been excluded from it.

Looking at the intersections, one can notice that more than half of the studies interested in bug data also consider source code.
That ratio goes up to more than 81\% for studies looking into commit metadata, which also consider source code.
Presumably, studies located in those intersections could not have been conducted by looking solely at non-source-code artifacts.
Once again, source code appears to be key for most ESE studies.

\smallskip
The general theme emerging from our answers to \Cref{rq:mined-artifacts,rq:artifact-combinations} appears to be that \textbf{while there is an increasing interest over time in mining novel software artifacts, it is complementing rather than replacing the need for mining source code}.

\subsection{\Cref{rq:artifact-purposes}: study purposes and software artifacts}

\begin{figure}
  \centering
  \includegraphics[width=0.8\columnwidth]{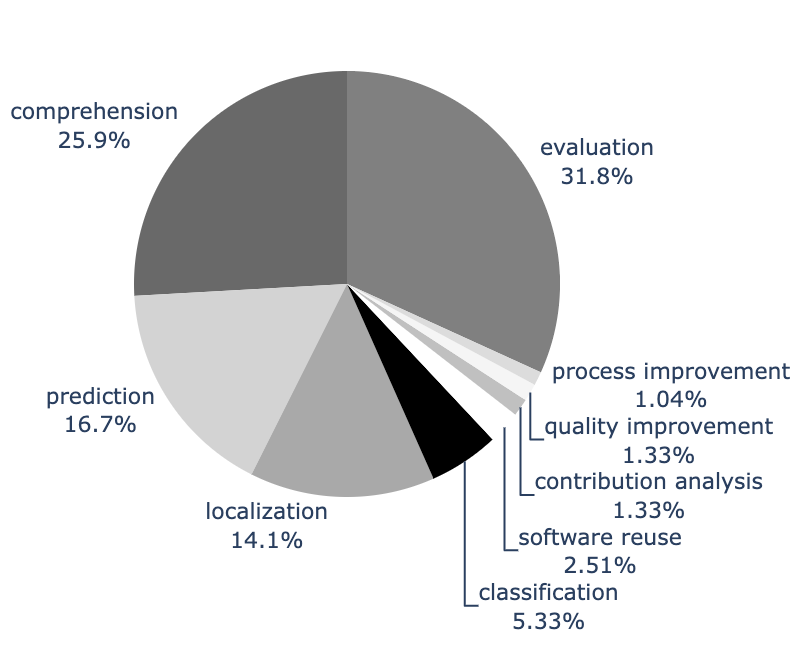}
  \caption{Ratio of papers pursuing a specific purpose.}
\label{fig:study-purposes}
\end{figure}

\begin{figure*}
  \centering
  \includegraphics[width=0.7\textwidth]{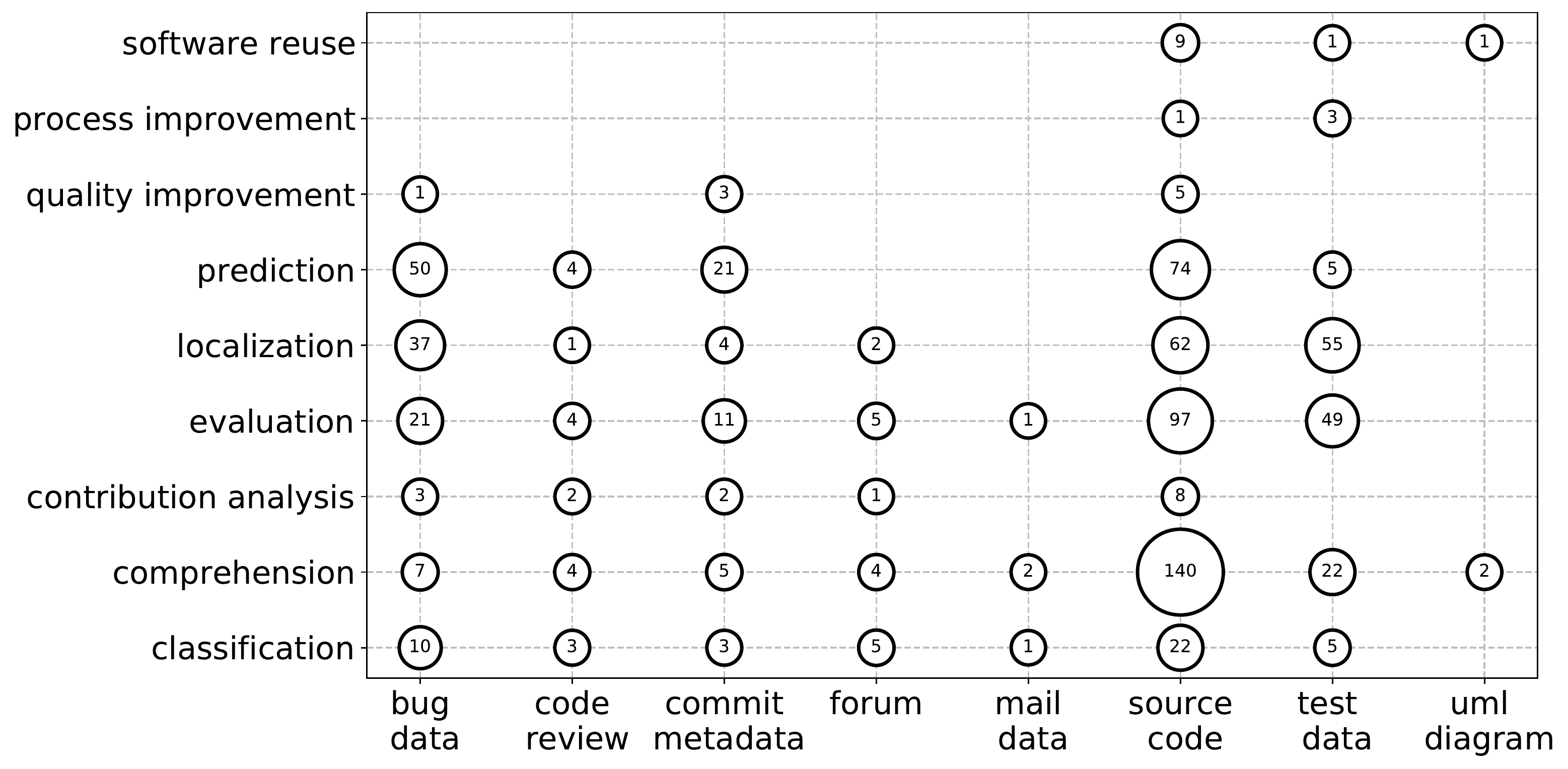}
  \caption{Study purposes v.~type of mined software artifacts, shown as a
    bubble chart for the entire corpus. Each bubble denotes the total number of
    papers detected as having a given purpose and mining a given type of
    artifact.}
  \label{fig:scatterPurpose}
\end{figure*}

\Cref{fig:study-purposes} shows a breakdown of papers mining software artifacts by detected study purpose, among those of \Cref{tab:purpose-taxonomy}.
The category “evaluation” is the highest represented with 1/3 of the studies, followed by “comprehension” (25.9\%).
``Prediction'' comes next (16.7\%), followed by ``localization'' (14.1\%).
The remaining five purposes represent in total less than 10\% of all studies. 

Overall, \textbf{ESE studies are most concerned with the evaluation of software systems}, followed by comprehension (most likely: \emph{code} comprehension, due to the prevalence of source code in our answers to previous RQs), and property prediction and localization (most likely: \emph{defect} prediction and \emph{bug} localization, due to the importance of bug data in previous results).

To study the relationship between study purposes and mined software artifacts, we show in \Cref{fig:scatterPurpose} a bubble chart of the two dimensions.
The size of each point in it indicates the number of studies with a given purpose that mine a given type of software artifacts.
Note that only papers for which we could detect at least one type of mined software artifacts are included in the chart.

The most relevant artifact/purpose combination is the use of source code to go after system comprehension.
This is consistent with the relevance, both in software engineering at large and in our paper corpus, of sub-fields and venues like program comprehension/ICPC.
Looking at both the artifact and purpose axes centered at this particular combination in \Cref{fig:scatterPurpose}, we observe that source code is referenced by papers with all purposes except process improvement and, orthogonally, that program comprehension is pursued referencing \emph{all} types of software artifacts.
It seems that \textbf{researchers are looking for all possible empirical signals in the long-running quest of understanding software systems}.
Conversely, source code is pervasively exploited to pursue most purposes in ESE research.

Outside of the over-represented column of source code, noteworthy columns are those of test data (used the most for localization and evaluation) and bug data (prediction, localization and, interestingly, evaluation, likely as a measure of \emph{software quality}).
Commit metadata are most used for prediction studies, likely to predict defects, considering that prediction is also the largest entry in the bug data column.

 \section{Discussion}
\label{sec:discussion}

\paragraph{Comparison with previous findings}

As discussed in \Cref{sec:relatedwork}, the most similar previous works to ours are \cite{farias2016msrmapping,demeyer2013happy}. 
Farias \etal~\cite{farias2016msrmapping} performed a systematic mapping study on 107 papers from the MSR conference.
They found that “comprehension \emph{of defects}” and  “[source] code” were the most common purpose and mined artifact, respectively.
They also found that structured artifacts were more mined than unstructured ones, but that the latter's use was increasing.
On a larger and more diverse corpus, we have confirmed that source code is the most mined artifact in empirical software engineering (ESE) research.
However, we find that comprehension \emph{of source code} (rather than defects) is the most common study purpose/artifact combination.

Demeyer \etal~\cite{demeyer2013happy}, based on an analysis of papers published at MSR, predicted that the trend to mine more and more unstructured artifacts would continue in the short term.
We have shown that this has not happened, neither at MSR nor across our full corpus.
We have also established that unstructured software artifacts are, for the most part, looked at in conjunction with structured ones.

In summary, our findings on a larger corpus of ESE papers only partly replicate previous findings on MSR papers~\cite{farias2016msrmapping,demeyer2013happy}.
MSR trends cannot be extrapolated to all ESE studies in our corpus, and some short-term trends there did not continue as initially predicted.\\

\textit{Implications of the findings.}
 Source code emerges as the most used and perennial artifact of interest in ESE research.
It is the top artifact across all analyzed conference papers, in almost every conference edition; it also supports almost all study purposes in the field.
As such, any research support initiative meant to either increase the availability of source code or facilitate mining is worth pursuing to help ESE researchers.
In this sense, the emergence of open source software in the 90s and collaborative social coding~\cite{dabbish2012socialcoding} in the 2010s have supported ESE research for more than a decade now.
Large-scale analysis platforms (e.g.,~\cite{dyer2013boa, swhipres2017, mockus2019woc}) and datasets (e.g., ~\cite{GHTorrent, swh-msr2019-dataset}) for both FOSS source code and related artifacts (commit metadata, bug, etc., which were all mined in the vast majority of analyzed papers) all go in the right direction to support future ESE research.
It would be interesting to investigate as future work if the availability of increasingly larger datasets and platforms is resulting in an actual increase of the size of the software artifact corpora that ESE researchers analyze, or if instead we still analyze same-sized ones (e.g., a relatively small sample of GitHub repositories, extrapolating from there).

The high reliance on software artifacts we have observed poses reliability challenges, particularly for replicability.
All analyzed artifacts must be (1) made available and (2) archived in the long-term for as long as replicability is pertinent (possibly forever).
Present-day accessibility of artifacts mined in ESE papers remains to be explored in future work, but the diversity of analyzed artifacts is enough to reveal a challenge: we need long-term platforms capable of both archiving and referencing with persistent identifiers a variety of artifact types, both structured and unstructured.
The cost of doing so will go up with the size of the analyzed corpora.

We can also deduct insights for producers of open datasets meant to be used in ESE research.
While the usefulness of such datasets has been established~\cite{kotti2020msrdatapapers}, based on the need to join together multiple types of artifacts for conducting ESE research which we have observed, we speculate that the importance of having mutually-consistent datasets will increase in the future.
For example, source code datasets that cannot be easily aligned with the corresponding bug and code review data will be less useful (and hence less impactful) than datasets that can.
Addressing this problem will require either making sure that ESE datasets provide stable identifiers that permit to easily cross-reference artifacts across heterogeneous datasets, or producing larger ``horizontal'' datasets that include all artifact types from the software systems of interest, e.g., all GitLab data for a given software ecosystem in a consistent snapshot, as opposed to a set of Git dumps and a separate database dump for social coding events.

 \section{Threats to Validity}
\label{sec:threats}

We discuss below threats to our findings validity, as well as applied mitigations, following the structure of Runeson and Hoes~\cite{runeson2009guidelines}.

\paragraph{Construct validity} 

The main threat to construct validity is that we categorize papers automatically based on n-gram occurrences rather than manually verifying all of them.
We have quantified the impact of this risk by manually verifying 200 papers, obtaining a satisfactory accuracy (see \Cref{sec:method}).
We have nevertheless taken steps to mitigate this threat: putting thresholds on the number of n-gram occurrences to avoid incidental references, as well as excluding paper sections where those references are common.
Relying on n-grams is also consistent with the state-of-the-art of meta-analyses in ESE~\cite{demeyer2013happy, amann2013software}.

In our n-gram analysis, we could have also have missed some n-grams that would denote mining of software artifacts of interest, leading to paper misclassification. We partly mitigated this by spot-checking papers for which we did not detect any artifact, iterating on the association between paper classes and n-grams in \Cref{tab:ngram-taxonomy,tab:purpose-taxonomy} until convergence.

\paragraph{Internal validity} 

We relied on CERMINE to extract textual content from PDF files.
The choice of this tool is based on a third-party quality assessment~\cite{tkaczyk2015cermine} pertinent to our use case and on specific functional requirements (e.g., we needed to break down papers by section) and non-functional requirements (e.g., we wanted the tool to be open source to ease replicability).
In addition to the documented \numNonParsablePapers papers we could not parse, conversion errors might have occurred and influenced obtained results.
We mitigated this risk by conducting spot checks on a random subset of papers in the corpus without noticing any relevant issues.

\paragraph{External validity}

We conducted a systematic mapping of papers mining software artifacts, without focusing on a specific venue.
We did, however, select both specific conferences and years (\Cref{tab:venues}), basing our choices on similar selections performed in related work and on an investigation of which venues welcome and/or regularly publish empirical software engineering (ESE).
We might have overlooked studies or entire venues that did not match our selection criteria but should have been included. We do not claim full generality of our findings for the ESE field.
But we expect them to capture software artifact mining trends and observe that our choices are consistent with (and our corpus generally larger than) analogous studies in the literature.

We relied on DBLP as ground truth for papers published in a given venue and year.
We could have missed papers that DBLP either lacks or has incorrectly indexed.
Given the preeminence of DBLP in computer science and that of its dataset for meta-analyses in the field, we consider this risk to be low.
Also, considering the large number of papers retained after filtering, we believe the results we report about to still be valid and of interest.

\paragraph{Reliability}

To mitigate reliability risks, we complement the description of our experimental methodology in \Cref{sec:method} with a complete replication package~\cite{replication-package}.
We encourage replication of our findings.

 \section{Conclusion}
\label{sec:conclusion}

\paragraph{Paper summary}

We have mined \numPapersIn papers from \numVenues top conferences that publish papers in the field of empirical software engineering (ESE) with NLP techniques, and n-gram analyses in particular, to map the use of software artifacts in the field over a period of \numYears years.
We map the most mined software artifacts in ESE studies, finding that the majority of papers in our corpus conspicuously mine at least one type of software artifacts.

Source code is stably the most frequently mined artifact, followed by test data; the use of other artifact types is more varied, both over time and across conferences.
The combined mining of different types of artifacts are significant: about 1/3 of all papers do that, with source code almost always being mined together with other artifacts.
Comparing study purposes with mined artifacts, we find that source code analysis supports all study purposes in ESE and that system comprehension is a major interest in the community, supported by the mining of all sorts of artifacts.
Our findings imply that increased care should be put into publishing open datasets that mix together different types of artifacts as consistent snapshots, and that digital preservation initiatives should diversify the type of artifacts they allow to archive and reference, in order to better support study replicability.

\paragraph{Future work}

Several directions remain to be explored as future work.
Our long-term goal with this work is to assess the degree of repeatability of ESE studies.
Establishing which types of artifacts are needed was a necessary first step, to be followed by an analysis of where artifacts come from (e.g., which collaborative coding platform) and an empirical assessment of their present-day availability---either as part of replication packages, at their original locations (if properly documented in studies), or from long-term digital archives.
We plan to explore the artifact origin and artifact availability dimensions next.

Second, we would like to characterize the scale at which ESE studies are conducted or, equivalently, the size of the datasets being used.
It is not clear neither if the increasing availability of larger and larger datasets and analysis platforms are really helping the community in conducting larger and larger empirical studies, nor if such studies are actually needed to move the state-of-the-art forward.
They are both important questions for the ESE community that we intend to answer next.

\clearpage

%%% -*-BibTeX-*-
%%% Do NOT edit. File created by BibTeX with style
%%% ACM-Reference-Format-Journals [18-Jan-2012].


%%% -*-BibTeX-*-
%%% Do NOT edit. File created by BibTeX with style
%%% ACM-Reference-Format-Journals [18-Jan-2012].

\begin{thebibliography}{34}

%%% ====================================================================
%%% NOTE TO THE USER: you can override these defaults by providing
%%% customized versions of any of these macros before the \bibliography
%%% command.  Each of them MUST provide its own final punctuation,
%%% except for \shownote{}, \showDOI{}, and \showURL{}.  The latter two
%%% do not use final punctuation, in order to avoid confusing it with
%%% the Web address.
%%%
%%% To suppress output of a particular field, define its macro to expand
%%% to an empty string, or better, \unskip, like this:
%%%
%%% \newcommand{\showDOI}[1]{\unskip}   % LaTeX syntax
%%%
%%% \def \showDOI #1{\unskip}           % plain TeX syntax
%%%
%%% ====================================================================

\ifx \showCODEN    \undefined \def \showCODEN     #1{\unskip}     \fi
\ifx \showDOI      \undefined \def \showDOI       #1{#1}\fi
\ifx \showISBNx    \undefined \def \showISBNx     #1{\unskip}     \fi
\ifx \showISBNxiii \undefined \def \showISBNxiii  #1{\unskip}     \fi
\ifx \showISSN     \undefined \def \showISSN      #1{\unskip}     \fi
\ifx \showLCCN     \undefined \def \showLCCN      #1{\unskip}     \fi
\ifx \shownote     \undefined \def \shownote      #1{#1}          \fi
\ifx \showarticletitle \undefined \def \showarticletitle #1{#1}   \fi
\ifx \showURL      \undefined \def \showURL       {\relax}        \fi
% The following commands are used for tagged output and should be
% invisible to TeX
\providecommand\bibfield[2]{#2}
\providecommand\bibinfo[2]{#2}
\providecommand\natexlab[1]{#1}
\providecommand\showeprint[2][]{arXiv:#2}

\bibitem[Abou~Khalil and Zacchiroli(2022)]%
        {replication-package}
\bibfield{author}{\bibinfo{person}{Zeinab Abou~Khalil} {and}
  \bibinfo{person}{Stefano Zacchiroli}.} \bibinfo{year}{2022}\natexlab{}.
\newblock \bibinfo{title}{Software Artifact Mining in Software Engineering
  Conferences: A Meta-Analysis --- Replication Package}.
\newblock \bibinfo{howpublished}{\url{https://doi.org/10.5281/zenodo.5877778}}.
\newblock


\bibitem[Amann et~al\mbox{.}(2014)]%
        {amann2013software}
\bibfield{author}{\bibinfo{person}{Sven Amann}, \bibinfo{person}{Stefanie
  Beyer}, \bibinfo{person}{Katja Kevic}, {and} \bibinfo{person}{Harald~C.
  Gall}.} \bibinfo{year}{2014}\natexlab{}.
\newblock \showarticletitle{Software Mining Studies: Goals, Approaches,
  Artifacts, and Replicability}. In \bibinfo{booktitle}{\emph{Software
  Engineering - International Summer Schools, {LASER} 2013-2014, Elba, Italy,
  Revised Tutorial Lectures}} \emph{(\bibinfo{series}{Lecture Notes in Computer
  Science}, Vol.~\bibinfo{volume}{8987})},
  \bibfield{editor}{\bibinfo{person}{Bertrand Meyer} {and}
  \bibinfo{person}{Martin Nordio}} (Eds.). \bibinfo{publisher}{Springer},
  \bibinfo{pages}{121--158}.
\newblock
\urldef\tempurl%
\url{https://doi.org/10.1007/978-3-319-28406-4\_5}
\showDOI{\tempurl}


\bibitem[Basili et~al\mbox{.}(1999)]%
        {basili1999building}
\bibfield{author}{\bibinfo{person}{Victor~R Basili}, \bibinfo{person}{Forrest
  Shull}, {and} \bibinfo{person}{Filippo Lanubile}.}
  \bibinfo{year}{1999}\natexlab{}.
\newblock \showarticletitle{Building knowledge through families of
  experiments}.
\newblock \bibinfo{journal}{\emph{IEEE Transactions on Software Engineering}}
  \bibinfo{volume}{25}, \bibinfo{number}{4} (\bibinfo{year}{1999}),
  \bibinfo{pages}{456--473}.
\newblock


\bibitem[Boisvert(2016)]%
        {boisvert2016acmbadges}
\bibfield{author}{\bibinfo{person}{Ronald~F. Boisvert}.}
  \bibinfo{year}{2016}\natexlab{}.
\newblock \showarticletitle{Incentivizing reproducibility}.
\newblock \bibinfo{journal}{\emph{Commun. {ACM}}} \bibinfo{volume}{59},
  \bibinfo{number}{10} (\bibinfo{year}{2016}), \bibinfo{pages}{5}.
\newblock
\urldef\tempurl%
\url{https://doi.org/10.1145/2994031}
\showDOI{\tempurl}


\bibitem[Dabbish et~al\mbox{.}(2012)]%
        {dabbish2012socialcoding}
\bibfield{author}{\bibinfo{person}{Laura Dabbish}, \bibinfo{person}{Colleen
  Stuart}, \bibinfo{person}{Jason Tsay}, {and} \bibinfo{person}{Jim Herbsleb}.}
  \bibinfo{year}{2012}\natexlab{}.
\newblock \showarticletitle{Social coding in GitHub: transparency and
  collaboration in an open software repository}. In
  \bibinfo{booktitle}{\emph{Proceedings of the ACM 2012 conference on computer
  supported cooperative work}}. ACM, \bibinfo{pages}{1277--1286}.
\newblock


\bibitem[de~Freitas~Farias et~al\mbox{.}(2016)]%
        {farias2016msrmapping}
\bibfield{author}{\bibinfo{person}{M{\'{a}}rio~Andr{\'{e}} de Freitas~Farias},
  \bibinfo{person}{Renato~Lima Novais}, \bibinfo{person}{Methanias~Cola{\c{c}}o
  J{\'{u}}nior}, \bibinfo{person}{Luis~Paulo da Silva~Carvalho},
  \bibinfo{person}{Manoel~G. Mendon{\c{c}}a}, {and}
  \bibinfo{person}{Rodrigo~Oliveira Sp{\'{\i}}nola}.}
  \bibinfo{year}{2016}\natexlab{}.
\newblock \showarticletitle{A systematic mapping study on mining software
  repositories}. In \bibinfo{booktitle}{\emph{Proceedings of the 31st Annual
  {ACM} Symposium on Applied Computing, Pisa, Italy, April 4-8, 2016}},
  \bibfield{editor}{\bibinfo{person}{Sascha Ossowski}} (Ed.).
  \bibinfo{publisher}{{ACM}}, \bibinfo{pages}{1472--1479}.
\newblock
\urldef\tempurl%
\url{https://doi.org/10.1145/2851613.2851786}
\showDOI{\tempurl}


\bibitem[Demeyer et~al\mbox{.}(2013)]%
        {demeyer2013happy}
\bibfield{author}{\bibinfo{person}{Serge Demeyer}, \bibinfo{person}{Alessandro
  Murgia}, \bibinfo{person}{Kevin Wyckmans}, {and} \bibinfo{person}{Ahmed
  Lamkanfi}.} \bibinfo{year}{2013}\natexlab{}.
\newblock \showarticletitle{Happy birthday! a trend analysis on past {MSR}
  papers}. In \bibinfo{booktitle}{\emph{Proceedings of the 10th Working
  Conference on Mining Software Repositories, {MSR} '13, San Francisco, CA,
  USA, May 18-19, 2013}}, \bibfield{editor}{\bibinfo{person}{Thomas
  Zimmermann}, \bibinfo{person}{Massimiliano~Di Penta}, {and}
  \bibinfo{person}{Sunghun Kim}} (Eds.). \bibinfo{publisher}{{IEEE} Computer
  Society}, \bibinfo{pages}{353--362}.
\newblock
\urldef\tempurl%
\url{https://doi.org/10.1109/MSR.2013.6624049}
\showDOI{\tempurl}


\bibitem[Di~Cosmo and Zacchiroli(2017)]%
        {swhipres2017}
\bibfield{author}{\bibinfo{person}{Roberto Di~Cosmo} {and}
  \bibinfo{person}{Stefano Zacchiroli}.} \bibinfo{year}{2017}\natexlab{}.
\newblock \showarticletitle{{Software Heritage}: Why and How to Preserve
  Software Source Code}. In \bibinfo{booktitle}{\emph{Proceedings of the 14th
  International Conference on Digital Preservation, iPRES 2017}}.
\newblock
\urldef\tempurl%
\url{https://hal.archives-ouvertes.fr/hal-01590958/}
\showURL{%
\tempurl}


\bibitem[Dyer et~al\mbox{.}(2013)]%
        {dyer2013boa}
\bibfield{author}{\bibinfo{person}{Robert Dyer}, \bibinfo{person}{Hoan~Anh
  Nguyen}, \bibinfo{person}{Hridesh Rajan}, {and} \bibinfo{person}{Tien~N
  Nguyen}.} \bibinfo{year}{2013}\natexlab{}.
\newblock \showarticletitle{Boa: A language and infrastructure for analyzing
  ultra-large-scale software repositories}. In
  \bibinfo{booktitle}{\emph{Proceedings of the 2013 International Conference on
  Software Engineering}}. IEEE Press, \bibinfo{pages}{422--431}.
\newblock


\bibitem[Felderer and Travassos(2020)]%
        {felderer2020esebook}
\bibfield{editor}{\bibinfo{person}{Michael Felderer} {and}
  \bibinfo{person}{Guilherme~Horta Travassos}} (Eds.).
  \bibinfo{year}{2020}\natexlab{}.
\newblock \bibinfo{booktitle}{\emph{Contemporary Empirical Methods in Software
  Engineering}}.
\newblock \bibinfo{publisher}{Springer}.
\newblock
\showISBNx{978-3-030-32488-9}
\urldef\tempurl%
\url{https://doi.org/10.1007/978-3-030-32489-6}
\showDOI{\tempurl}


\bibitem[Felizardo and Carver(2020)]%
        {felizardo2020automating}
\bibfield{author}{\bibinfo{person}{Katia~R Felizardo} {and}
  \bibinfo{person}{Jeffrey~C Carver}.} \bibinfo{year}{2020}\natexlab{}.
\newblock \showarticletitle{Automating systematic literature review}.
\newblock \bibinfo{journal}{\emph{Contemporary Empirical Methods in Software
  Engineering}} (\bibinfo{year}{2020}), \bibinfo{pages}{327--355}.
\newblock


\bibitem[Gousios and Spinellis(2012)]%
        {GHTorrent}
\bibfield{author}{\bibinfo{person}{Georgios Gousios} {and}
  \bibinfo{person}{Diomidis Spinellis}.} \bibinfo{year}{2012}\natexlab{}.
\newblock \showarticletitle{GHTorrent: Github's data from a firehose}. In
  \bibinfo{booktitle}{\emph{9th {IEEE} Working Conference of Mining Software
  Repositories, {MSR}}}, \bibfield{editor}{\bibinfo{person}{Michele Lanza},
  \bibinfo{person}{Massimiliano~Di Penta}, {and} \bibinfo{person}{Tao Xie}}
  (Eds.). \bibinfo{publisher}{{IEEE} Computer Society},
  \bibinfo{pages}{12--21}.
\newblock
\showISBNx{978-1-4673-1761-0}
\urldef\tempurl%
\url{https://doi.org/10.1109/MSR.2012.6224294}
\showDOI{\tempurl}


\bibitem[Hassan(2008)]%
        {hassan2008road}
\bibfield{author}{\bibinfo{person}{Ahmed~E Hassan}.}
  \bibinfo{year}{2008}\natexlab{}.
\newblock \showarticletitle{The road ahead for mining software repositories}.
  In \bibinfo{booktitle}{\emph{2008 Frontiers of Software Maintenance}}. IEEE,
  \bibinfo{pages}{48--57}.
\newblock


\bibitem[Hemmati et~al\mbox{.}(2013)]%
        {hemmati2013msr}
\bibfield{author}{\bibinfo{person}{Hadi Hemmati}, \bibinfo{person}{Sarah Nadi},
  \bibinfo{person}{Olga Baysal}, \bibinfo{person}{Oleksii Kononenko},
  \bibinfo{person}{Wei Wang}, \bibinfo{person}{Reid Holmes}, {and}
  \bibinfo{person}{Michael~W Godfrey}.} \bibinfo{year}{2013}\natexlab{}.
\newblock \showarticletitle{The {MSR} cookbook: Mining a decade of research}.
  In \bibinfo{booktitle}{\emph{2013 10th Working Conference on Mining Software
  Repositories (MSR)}}. IEEE, \bibinfo{pages}{343--352}.
\newblock


\bibitem[Howison et~al\mbox{.}(2006)]%
        {FLOSSmole}
\bibfield{author}{\bibinfo{person}{James Howison}, \bibinfo{person}{Megan
  Conklin}, {and} \bibinfo{person}{Kevin Crowston}.}
  \bibinfo{year}{2006}\natexlab{}.
\newblock \showarticletitle{FLOSSmole: {A} Collaborative Repository for {FLOSS}
  Research Data and Analyses}.
\newblock \bibinfo{journal}{\emph{{IJITWE}}} \bibinfo{volume}{1},
  \bibinfo{number}{3} (\bibinfo{year}{2006}), \bibinfo{pages}{17--26}.
\newblock
\urldef\tempurl%
\url{https://doi.org/10.4018/jitwe.2006070102}
\showDOI{\tempurl}


\bibitem[Ioannidis(2010)]%
        {ioannidis2010metaresearch}
\bibfield{author}{\bibinfo{person}{John P.~A. Ioannidis}.}
  \bibinfo{year}{2010}\natexlab{}.
\newblock \showarticletitle{Meta-research: The art of getting it wrong}.
\newblock \bibinfo{journal}{\emph{Res. Synth. Methods}} \bibinfo{volume}{1},
  \bibinfo{number}{3-4} (\bibinfo{date}{Jul} \bibinfo{year}{2010}),
  \bibinfo{pages}{169--184}.
\newblock
\showISSN{1759-2879}
\urldef\tempurl%
\url{https://doi.org/10.1002/jrsm.19}
\showDOI{\tempurl}


\bibitem[Kagdi et~al\mbox{.}(2007)]%
        {kagdi2007survey}
\bibfield{author}{\bibinfo{person}{Huzefa Kagdi}, \bibinfo{person}{Michael~L
  Collard}, {and} \bibinfo{person}{Jonathan~I Maletic}.}
  \bibinfo{year}{2007}\natexlab{}.
\newblock \showarticletitle{A survey and taxonomy of approaches for mining
  software repositories in the context of software evolution}.
\newblock \bibinfo{journal}{\emph{Journal of software maintenance and
  evolution: Research and practice}} \bibinfo{volume}{19}, \bibinfo{number}{2}
  (\bibinfo{year}{2007}), \bibinfo{pages}{77--131}.
\newblock


\bibitem[Kettunen et~al\mbox{.}(2005)]%
        {kettunen2005stem}
\bibfield{author}{\bibinfo{person}{Kimmo Kettunen}, \bibinfo{person}{Tuomas
  Kunttu}, {and} \bibinfo{person}{Kalervo J{\"a}rvelin}.}
  \bibinfo{year}{2005}\natexlab{}.
\newblock \showarticletitle{To stem or lemmatize a highly inflectional language
  in a probabilistic {IR} environment?}
\newblock \bibinfo{journal}{\emph{Journal of Documentation}}
  (\bibinfo{year}{2005}).
\newblock


\bibitem[Kotti et~al\mbox{.}(2020)]%
        {kotti2020msrdatapapers}
\bibfield{author}{\bibinfo{person}{Zoe Kotti}, \bibinfo{person}{Konstantinos
  Kravvaritis}, \bibinfo{person}{Konstantina Dritsa}, {and}
  \bibinfo{person}{Diomidis Spinellis}.} \bibinfo{year}{2020}\natexlab{}.
\newblock \showarticletitle{Standing on shoulders or feet? An extended study on
  the usage of the {MSR} data papers}.
\newblock \bibinfo{journal}{\emph{Empir. Softw. Eng.}} \bibinfo{volume}{25},
  \bibinfo{number}{5} (\bibinfo{year}{2020}), \bibinfo{pages}{3288--3322}.
\newblock
\urldef\tempurl%
\url{https://doi.org/10.1007/s10664-020-09834-7}
\showDOI{\tempurl}


\bibitem[Ley(2002)]%
        {ley2002dblp}
\bibfield{author}{\bibinfo{person}{Michael Ley}.}
  \bibinfo{year}{2002}\natexlab{}.
\newblock \showarticletitle{The {DBLP} Computer Science Bibliography:
  Evolution, Research Issues, Perspectives}. In
  \bibinfo{booktitle}{\emph{String Processing and Information Retrieval, 9th
  International Symposium, {SPIRE} 2002, Lisbon, Portugal, September 11-13,
  2002, Proceedings}} \emph{(\bibinfo{series}{Lecture Notes in Computer
  Science}, Vol.~\bibinfo{volume}{2476})},
  \bibfield{editor}{\bibinfo{person}{Alberto H.~F. Laender} {and}
  \bibinfo{person}{Arlindo~L. Oliveira}} (Eds.). \bibinfo{publisher}{Springer},
  \bibinfo{pages}{1--10}.
\newblock
\urldef\tempurl%
\url{https://doi.org/10.1007/3-540-45735-6\_1}
\showDOI{\tempurl}


\bibitem[Ma et~al\mbox{.}(2019)]%
        {mockus2019woc}
\bibfield{author}{\bibinfo{person}{Yuxing Ma}, \bibinfo{person}{Chris Bogart},
  \bibinfo{person}{Sadika Amreen}, \bibinfo{person}{Russell Zaretzki}, {and}
  \bibinfo{person}{Audris Mockus}.} \bibinfo{year}{2019}\natexlab{}.
\newblock \showarticletitle{World of code: an infrastructure for mining the
  universe of open source VCS data}. In \bibinfo{booktitle}{\emph{Proceedings
  of the 16th International Conference on Mining Software Repositories}}. IEEE
  Press, \bibinfo{pages}{143--154}.
\newblock


\bibitem[Manning and Schutze(1999)]%
        {manning1999foundations}
\bibfield{author}{\bibinfo{person}{Christopher Manning} {and}
  \bibinfo{person}{Hinrich Schutze}.} \bibinfo{year}{1999}\natexlab{}.
\newblock \bibinfo{booktitle}{\emph{Foundations of statistical natural language
  processing}}.
\newblock \bibinfo{publisher}{MIT press}.
\newblock


\bibitem[Mathew et~al\mbox{.}(2018)]%
        {menzies2018swetrends}
\bibfield{author}{\bibinfo{person}{George Mathew}, \bibinfo{person}{Amritanshu
  Agrawal}, {and} \bibinfo{person}{Tim Menzies}.}
  \bibinfo{year}{2018}\natexlab{}.
\newblock \showarticletitle{Finding Trends in Software Research}.
\newblock \bibinfo{journal}{\emph{{IEEE} Transactions on Software Engineering}}
  (\bibinfo{year}{2018}).
\newblock
\urldef\tempurl%
\url{https://doi.org/10.1109/TSE.2018.2870388}
\showDOI{\tempurl}
\newblock
\shownote{To appear}.


\bibitem[Moro et~al\mbox{.}(2015)]%
        {moro2015business}
\bibfield{author}{\bibinfo{person}{S{\'e}rgio Moro}, \bibinfo{person}{Paulo
  Cortez}, {and} \bibinfo{person}{Paulo Rita}.}
  \bibinfo{year}{2015}\natexlab{}.
\newblock \showarticletitle{Business intelligence in banking: A literature
  analysis from 2002 to 2013 using text mining and latent Dirichlet
  allocation}.
\newblock \bibinfo{journal}{\emph{Expert Systems with Applications}}
  \bibinfo{volume}{42}, \bibinfo{number}{3} (\bibinfo{year}{2015}),
  \bibinfo{pages}{1314--1324}.
\newblock


\bibitem[Nagappan et~al\mbox{.}(2009)]%
        {nagappan2009msr}
\bibfield{author}{\bibinfo{person}{Nachiappan Nagappan},
  \bibinfo{person}{Andreas Zeller}, {and} \bibinfo{person}{Thomas Zimmermann}.}
  \bibinfo{year}{2009}\natexlab{}.
\newblock \showarticletitle{Guest Editors' Introduction: Mining Software
  Archives}.
\newblock \bibinfo{journal}{\emph{{IEEE} Softw.}} \bibinfo{volume}{26},
  \bibinfo{number}{1} (\bibinfo{year}{2009}), \bibinfo{pages}{24--25}.
\newblock
\urldef\tempurl%
\url{https://doi.org/10.1109/MS.2009.14}
\showDOI{\tempurl}


\bibitem[Novais et~al\mbox{.}(2013)]%
        {novais2013software}
\bibfield{author}{\bibinfo{person}{Renato~Lima Novais},
  \bibinfo{person}{Andr{\'e} Torres}, \bibinfo{person}{Thiago~Souto Mendes},
  \bibinfo{person}{Manoel Mendon{\c{c}}a}, {and} \bibinfo{person}{Nico
  Zazworka}.} \bibinfo{year}{2013}\natexlab{}.
\newblock \showarticletitle{Software evolution visualization: A systematic
  mapping study}.
\newblock \bibinfo{journal}{\emph{Information and Software Technology}}
  \bibinfo{volume}{55}, \bibinfo{number}{11} (\bibinfo{year}{2013}),
  \bibinfo{pages}{1860--1883}.
\newblock


\bibitem[Pietri et~al\mbox{.}(2019)]%
        {swh-msr2019-dataset}
\bibfield{author}{\bibinfo{person}{Antoine Pietri}, \bibinfo{person}{Diomidis
  Spinellis}, {and} \bibinfo{person}{Stefano Zacchiroli}.}
  \bibinfo{year}{2019}\natexlab{}.
\newblock \showarticletitle{The {S}oftware {H}eritage graph dataset: public
  software development under one roof}. In
  \bibinfo{booktitle}{\emph{Proceedings of the 16th International Conference on
  Mining Software Repositories, {MSR} 2019, 26-27 May 2019, Montreal,
  Canada.}}, \bibfield{editor}{\bibinfo{person}{Margaret{-}Anne~D. Storey},
  \bibinfo{person}{Bram Adams}, {and} \bibinfo{person}{Sonia Haiduc}} (Eds.).
  \bibinfo{publisher}{{IEEE} / {ACM}}, \bibinfo{pages}{138--142}.
\newblock
\showISBNx{978-1-7281-3412-3}
\urldef\tempurl%
\url{https://dl.acm.org/citation.cfm?id=3341907}
\showURL{%
\tempurl}


\bibitem[Robles(2010)]%
        {robles2010replicating}
\bibfield{author}{\bibinfo{person}{Gregorio Robles}.}
  \bibinfo{year}{2010}\natexlab{}.
\newblock \showarticletitle{Replicating {MSR}: A study of the potential
  replicability of papers published in the mining software repositories
  proceedings}. In \bibinfo{booktitle}{\emph{2010 7th IEEE Working Conference
  on Mining Software Repositories (MSR 2010)}}. IEEE,
  \bibinfo{pages}{171--180}.
\newblock


\bibitem[Runeson and H{\"o}st(2009)]%
        {runeson2009guidelines}
\bibfield{author}{\bibinfo{person}{Per Runeson} {and} \bibinfo{person}{Martin
  H{\"o}st}.} \bibinfo{year}{2009}\natexlab{}.
\newblock \showarticletitle{Guidelines for conducting and reporting case study
  research in software engineering}.
\newblock \bibinfo{journal}{\emph{Empirical Software Engineering}}
  \bibinfo{volume}{14}, \bibinfo{number}{2} (\bibinfo{year}{2009}),
  \bibinfo{pages}{131}.
\newblock


\bibitem[Shull et~al\mbox{.}(2008)]%
        {shull2008eseguide}
\bibfield{editor}{\bibinfo{person}{Forrest Shull}, \bibinfo{person}{Janice
  Singer}, {and} \bibinfo{person}{Dag I.~K. Sj{\o}berg}} (Eds.).
  \bibinfo{year}{2008}\natexlab{}.
\newblock \bibinfo{booktitle}{\emph{Guide to Advanced Empirical Software
  Engineering}}.
\newblock \bibinfo{publisher}{Springer}.
\newblock
\showISBNx{9781848000438}
\urldef\tempurl%
\url{https://doi.org/10.1007/978-1-84800-044-5}
\showDOI{\tempurl}


\bibitem[Soper and Turel(2012)]%
        {soper2012n}
\bibfield{author}{\bibinfo{person}{Daniel~S Soper} {and} \bibinfo{person}{Ofir
  Turel}.} \bibinfo{year}{2012}\natexlab{}.
\newblock \showarticletitle{An n-gram analysis of Communications 2000--2010}.
\newblock \bibinfo{journal}{\emph{Commun. ACM}} \bibinfo{volume}{55},
  \bibinfo{number}{5} (\bibinfo{year}{2012}), \bibinfo{pages}{81--87}.
\newblock


\bibitem[Tkaczyk et~al\mbox{.}(2018)]%
        {tkaczyk2018machine}
\bibfield{author}{\bibinfo{person}{Dominika Tkaczyk}, \bibinfo{person}{Andrew
  Collins}, \bibinfo{person}{Paraic Sheridan}, {and} \bibinfo{person}{Joeran
  Beel}.} \bibinfo{year}{2018}\natexlab{}.
\newblock \showarticletitle{Machine learning vs. rules and out-of-the-box vs.
  retrained: An evaluation of open-source bibliographic reference and citation
  parsers}. In \bibinfo{booktitle}{\emph{Proceedings of the 18th ACM/IEEE on
  joint conference on digital libraries}}. \bibinfo{pages}{99--108}.
\newblock


\bibitem[Tkaczyk et~al\mbox{.}(2015)]%
        {tkaczyk2015cermine}
\bibfield{author}{\bibinfo{person}{Dominika Tkaczyk}, \bibinfo{person}{Pawel
  Szostek}, \bibinfo{person}{Mateusz Fedoryszak}, \bibinfo{person}{Piotr~Jan
  Dendek}, {and} \bibinfo{person}{Lukasz Bolikowski}.}
  \bibinfo{year}{2015}\natexlab{}.
\newblock \showarticletitle{{CERMINE:} automatic extraction of structured
  metadata from scientific literature}.
\newblock \bibinfo{journal}{\emph{Int. J. Document Anal. Recognit.}}
  \bibinfo{volume}{18}, \bibinfo{number}{4} (\bibinfo{year}{2015}),
  \bibinfo{pages}{317--335}.
\newblock
\urldef\tempurl%
\url{https://doi.org/10.1007/s10032-015-0249-8}
\showDOI{\tempurl}


\bibitem[Vasilescu et~al\mbox{.}(2013)]%
        {vasilescu2013sweconf}
\bibfield{author}{\bibinfo{person}{Bogdan Vasilescu},
  \bibinfo{person}{Alexander Serebrenik}, {and} \bibinfo{person}{Tom Mens}.}
  \bibinfo{year}{2013}\natexlab{}.
\newblock \showarticletitle{A historical dataset of software engineering
  conferences}. In \bibinfo{booktitle}{\emph{2013 10th Working Conference on
  Mining Software Repositories (MSR)}}. IEEE, \bibinfo{pages}{373--376}.
\newblock


\end{thebibliography}
\end{document}